\documentclass[%
 reprint,
superscriptaddress,
nofootinbib,
 amsmath,amssymb,
 aps,
 longbibliography,
floats,
floatfix,
]{revtex4-1}

\usepackage[utf8]{inputenc} % allow utf-8 input
\usepackage[T1]{fontenc}    % use 8-bit T1 fonts
\usepackage[colorlinks=true,urlcolor=blue,linkcolor=blue,citecolor=blue]{hyperref}       % hyperlinks
\usepackage{url}            % simple URL typesetting
\usepackage{booktabs}       % professional-quality tables
\usepackage{amsfonts}       % blackboard math symbols
\usepackage{nicefrac}       % compact symbols for 1/2, etc.
\usepackage{microtype}      % microtypography
\usepackage{natbib}
\usepackage{textgreek}
\usepackage{graphicx,caption}
\usepackage{wrapfig}
\usepackage[toc,page]{appendix}
\usepackage{subcaption}
\usepackage{todonotes}
\usepackage{afterpage}

\usepackage{multirow}
\usepackage{siunitx}

\usepackage{float}
\makeatletter
\let\newfloat\newfloat@ltx
\makeatother

\usepackage{algorithm}% http://ctan.org/pkg/algorithms
\usepackage{algpseudocode}% http://ctan.org/pkg/algorithmicx

\newcommand{\Sec}[1]{Section~#1}

\usepackage{algorithm}% http://ctan.org/pkg/algorithms
\usepackage{algpseudocode}% http://ctan.org/pkg/algorithmicx
\begin{document}

\preprint{APS/123-QED}

\title{Finetuning Foundation Models for Joint Analysis Optimization}

\author{Matthias Vigl}
\author{Nicole Hartman}
\author{Lukas Heinrich}
\affiliation{Technical University of Munich\\
Email: \texttt{matthias.vigl@tum.de}
}

\begin{abstract}
    In this work we demonstrate that significant gains in performance and data efficiency can be achieved in High Energy Physics (HEP) by moving beyond the standard paradigm of sequential optimization or reconstruction and analysis components. We conceptually connect HEP reconstruction and analysis to modern machine learning workflows such as pretraining, finetuning, domain adaptation and high-dimensional embedding spaces and quantify the gains in the example usecase of searches of heavy resonances decaying via an intermediate di-Higgs system to four $b$-jets.
\end{abstract}

\maketitle

\section{Introduction}
\label{sec:intro}

Data analysis in High Energy Physics (HEP) aims to make inferences on fundamental theories of nature based on data recorded at large-scale experiments, such as those at the Large Hadron Collider (LHC). The observed data at such experiments originates from high energy collisions and their evolution is modeled by a deep hierarchy of physical models, describing e.g. the decay of particles, their subsequent radiation patterns and finally the interactions with the detecting instrument. Consequently, the primary approach in data analysis is that of hierarchical pattern recognition and inference: first, low-level patterns in the detector data are identified and used to reconstruct properties of particles that directly interacted with the detector. Based on these, the earlier stages of the data-generating process are reconstructed in a hierarchical fashion before inferences on the originating theory can finally be made. That is, the inference pipeline aims to approximately \emph{invert} the data-generating process by progressively summarizing the data, reconstructing earlier latent states and subsequently analyzing those. Traditionally, the individual reconstruction and analysis algorithms are optimized sequentially (greedily), with late-stage algorithms being optimized on inputs of previously optimized earlier stages. While practical,  it is unlikely that this strategy would yield the \emph{jointly optimal} data analysis pipeline.

\begin{figure}[h]
    \centering
    \includegraphics[width=0.36\textwidth]{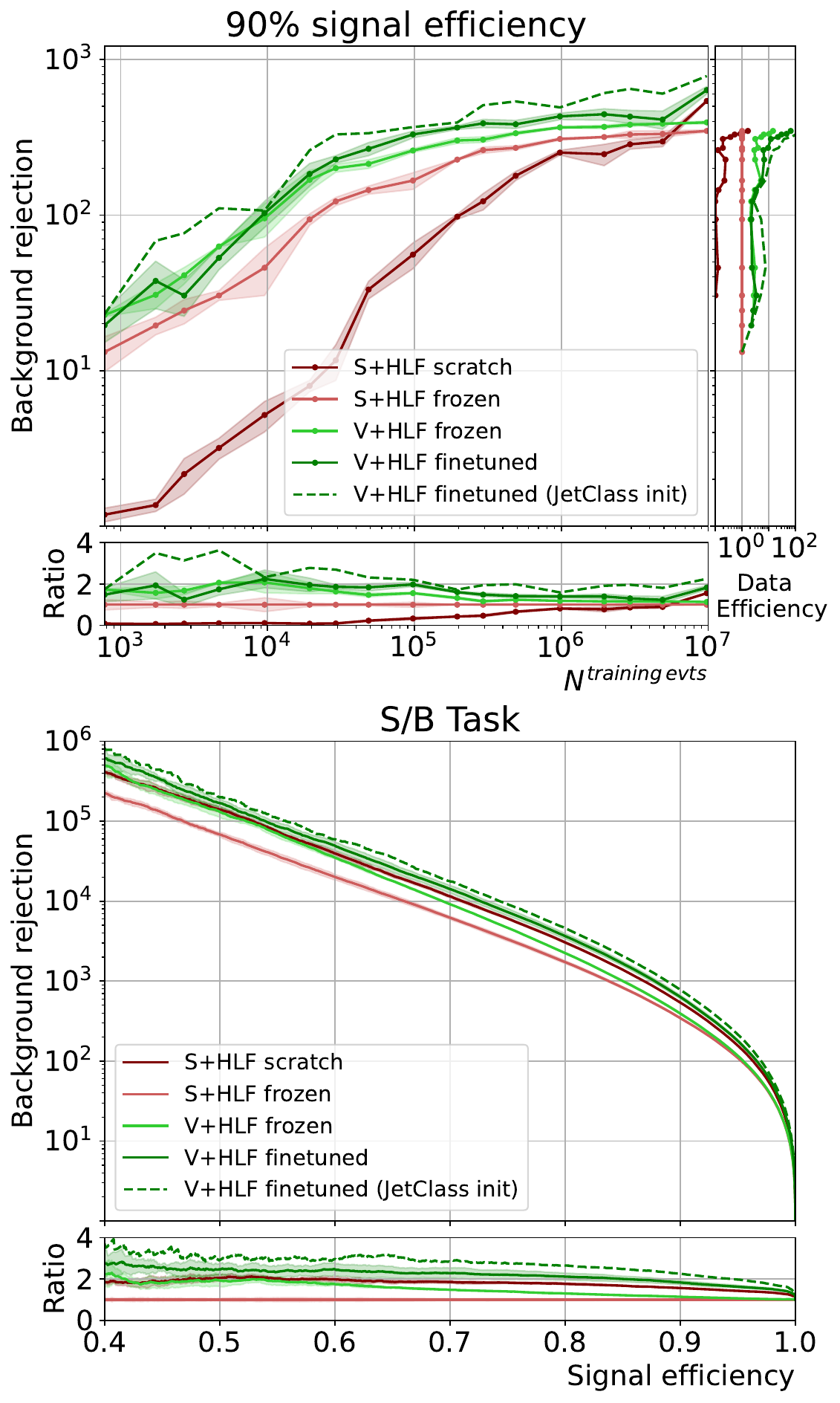}
    \caption{Strategies from modern machine learning such as finetuning, large-scale pretraining, finetuning, domain adaptation and high-dimensional embeddings (green curves) can lead to significant performance gains over the traditional HEP approach, denoted here as \texttt{S+HLF(frozen)}. Top: Performance evolution as a function of training dataset size. Bottom: Final Performance at 10M training samples.}
    \label{fig:fig1}
\end{figure}

In this work, we show that significant gains in performance and data efficiency can be achieved by instead pursuing a more global gradient-based optimization strategy and modelling the data analysis approach after modern large-scale machine learning (ML) workflows with foundation models. As shown in Figure~\ref{fig:fig1} these gains materialize as boosted performance at a fixed dataset size as well as an improved data efficiency, i.e. samples required to reach a desired level of performance.
This paper is outlined as follows:
In \Sec{\ref{sec:prior work}} we review relevant related work. In \Sec{\ref{sec:background}} we recall preliminaries from simulation-based inference and point out similarities between machine-learning with foundation models and common practice in particle physics. 
\Sec{\ref{sec:testcase_and_data}} introduces a demonstrator use-case for end-to-end optimization and discusses the datasets involved, whereas \Sec{\ref{sec:model}} discusses the neural network architectures and training strategies considered in the study. In
\Sec{\ref{sec:results}} we discuss the results while giving an outlook towards future research directions  in \Sec{\ref{sec:conclusions}}. Our main contributions are:
\begin{itemize}
    \item We establish a correspondence between concepts in the HEP analysis workflow and those in modern deep learning such as foundation models, downstream tasks and finetuning to describe a general strategy for optimizing HEP data analysis pipelines.
    \item We demonstrate, to our knowledge for the first time, a finetuning workflow in the hierarchical setting of per-object representation and event-level inference within particle physics.
    \item We quantify the significant gains due to end-to-end optimization with respect to data efficiency and performance at fixed sample size.
    \item We provide evidence of successful domain adaptation in a hierarchical setting of HEP foundation models finetuned on datasets other than the one they are pretrained with.
\end{itemize}

\section{Related Work}
\label{sec:prior work}
This work connects to a larger body of research concerned with the optimization of HEP analysis and the role of processing low-level variables with deep-learning systems~\cite{doi:10.1142/9789811234033_0003,doi:10.1142/9789811234033_0010}. Early work on neural networks with inductive bias informed by quantum chromodynamics~\cite{Louppe:2017ipp} investigated a hierarchical approach that jointly optimized a pipeline consisting of a neural embeddings of jets followed by an event classification but has not in detail studied performance under various pretraining strategies. Increasingly, hierarchies of neural networks algorithms are used within reconstruction for larger overall tasks, such as tracking~\cite{doi:10.1142/9789811234033_0012,Liu:2023siw,DeZoort:2021rbj} or particle flow reconstruction~\cite{Pata_2021,Di_Bello_2023}. However, they are often greedily optimized due to non-differentiable elements in the pipeline. To bridge this gap, approaches that enable gradient information to flow freely have grown into the rich research domain of differentiable programming, with e.g. differentiable vertexing~\cite{Smith:2023ssh}, statistical inference~\cite{DeCastro:2018psv,Simpson:2022suz,pyhf,pyhf_joss}, branching processes~\cite{Kagan:2023gxz,Nachman:2022jbj},  matrix-elements~\cite{Heinrich:2022xfa} or even detector-design~\cite{MODE:2022znx}. 

This work relies heavily on jet-level backbone models, which are primarily developed in the context of jet-tagging tasks~\cite{Kasieczka:2019dbj,Komiske:2018cqr}. Specifically we use the transformer-based \texttt{ParT}~\cite{Qu:2022mxj} as a jet representation backbone, but the method can be extended to other jet-level models that access the full low-level constituent data, such as \texttt{JetCLR}~\cite{Dillon:2021gag}, \texttt{LorentzNet}~\cite{Gong:2022lye}, or \texttt{GN2X}~\cite{ATL-PHYS-PUB-2023-021}.

The notion of general-purpose foundation models that are pretrained and then finetuned is commonplace in computer vision~\cite{bao2022beit,2023arXiv230407193O,2021arXiv210504906B} and natural language processing~\cite{lewis2019bart,devlin2019bert,openai2023gpt4,NEURIPS2020_1457c0d6}. Often, such foundation models aim to develop a self-supervised pre-training strategy; however, supervised strategies are also common~\cite{2020arXiv201011929D}. Increasingly, there are also efforts within the natural sciences to train and exploit general-purpose foundation models~\cite{ScaifeFoundation,2023arXiv231003024L,2023arXiv230309949S,2023arXiv230615794N}. Domain adaptation has been investigated previously in high-energy physics in a jet-tagging contexts~\cite{Qu:2022mxj,Dreyer:2022yom} but to our knowledge not in hierarchical configurations. In parallel to the present effort on supervised backbones, investigations are ongoing on the potential of \emph{self-supervised} backbones in HEP through \emph{masked particle modelling}, which extends the masked language modelling approach from NLP to the HEP domain~\cite{mpm}.

\section{Background}
\label{sec:background}

\begin{figure}
    \centering
    \includegraphics[width=0.45\textwidth]{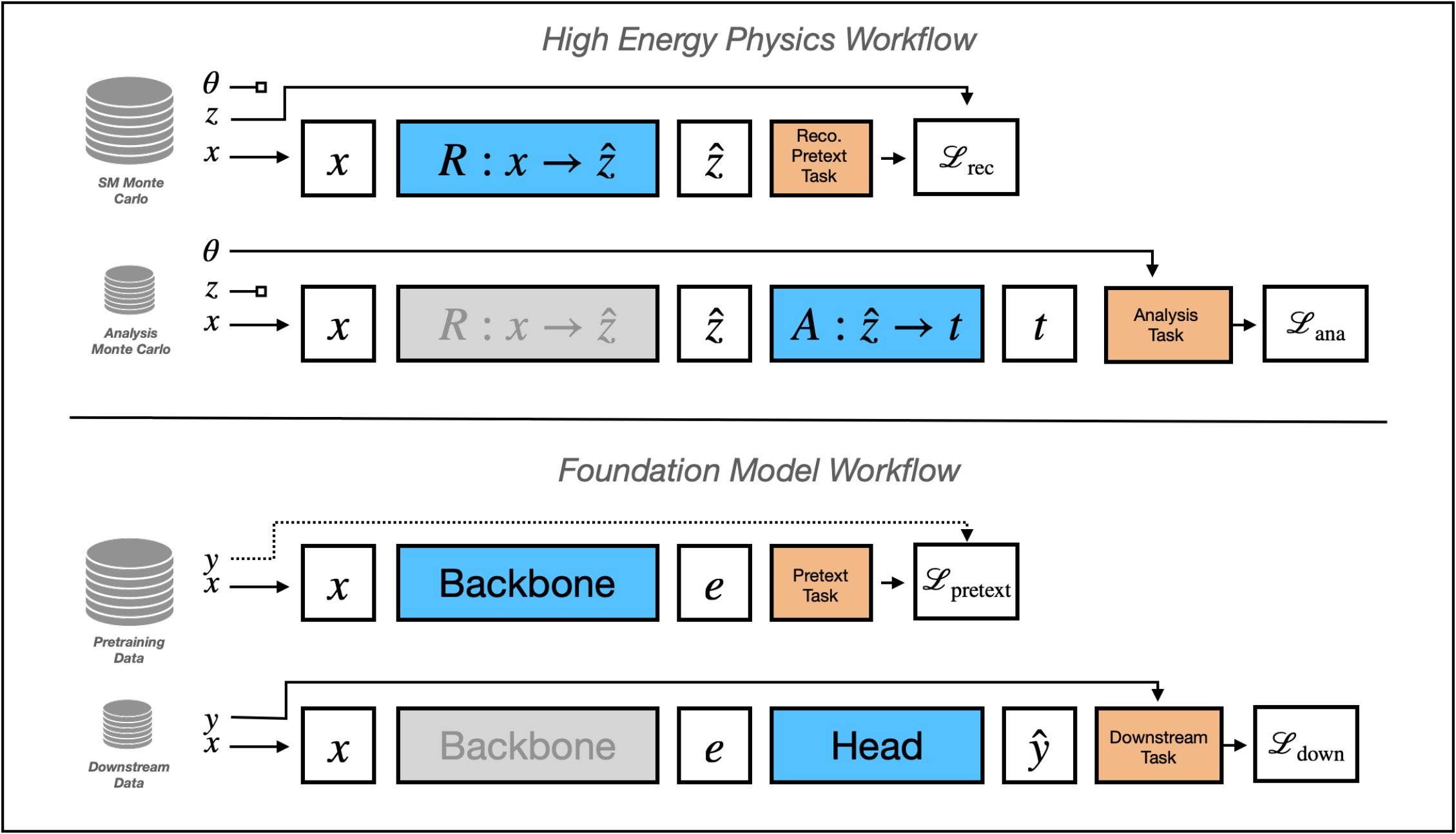}
    \caption{Modern machine learning and HEP data analysis exhibit conceptual similarities. Reconstruction plays the role of a backbone or foundation model yielding a general purpose representation of high-dimensional low-level data. The physics data analysis itself is a ``head'' that produces task-specific summary statistics.}
    \label{fig:FM_HEP}
\end{figure}

\subsection{Simulation-based Inference and Summary Statistics}

The data analysis strategy described in \Sec{\ref{sec:intro}}  can be motivated and formalized through the lens of simulation-based (or likelihood-free) inference~\cite{2020PNAS..11730055C}. In HEP, the evaluation of the likelihood $p(x|\theta)$ of the observed data $x$ given a theory $\theta$ is intractable due to the fact that the data-generating process proceeds through complex intermediate states that are not directly observed, such as particles decays, radiation effects and interactions with dense detector material. Formally, we can collect all such unobserved states into a single latent variable $z$. The likelihood-free nature then becomes apparent, as the evaluation of the likelihood would require the computation of a high-dimensional integral $p(x|\theta) = \int_z p(x|z)p(z|\theta)$. Inference in this setting is primarily enabled by the existence of high-quality simulators that encode the physics of the data-generating process, so that it's possible to obtain \emph{joint samples} $(x,z,\theta)\sim p(x|z)p(z|\theta)p(\theta)$ through ancestral sampling. A direct density estimation of $p(x|\theta)$ based on the resulting marginal samples $x\sim p(x|\theta)$ is however impossible due to the high dimensionality of the data $x$, which denotes the readouts of $O(10^8)$ sensors of modern  physics experiments such as those at the LHC.

The dominant method to perform inference on the theory parameters $\theta$ is therefore through the density estimation of suitable low-dimensional  \emph{summary statistics} $T: x \mapsto t$ followed by standard statistical inference techniques. The computation of the summary statistic is often conceptually split into a \emph{reconstruction-level summary} and \emph{analysis-level summary}~\footnote{The exact delineation of where reconstruction ends and analysis begins is a matter of interpretation.}. Formally, we can state that the goal of reconstruction is to map the low-level data $x$ into an event record representation, i.e. an estimate $\hat{z}$ of the latent state in the form of lists of particles in the event and their properties. The reconstruction of the data analysis is generally thought of as a generic preprocessing step that already drastically reduces the dimensionality of the data and is highly interpretable. While it is important to note that reconstruction is not a monolithic neural network, but rather a complex composite of both non-neural and neural components, for the purposes of this discussion it can be thought of as a single parametrized function $R_\rho: x \mapsto \hat{z}$, where $\rho$ stands for variable parameters that control the details of the reconstruction process.

The reconstruction phase is then followed by a more case-dependent \emph{analysis phase} that drives the final inference. Again setting aside many important details of HEP analysis, we can define as its core the definition of a task-dependent summary statistic $A_\alpha: \hat{z} \to t$, where $\alpha$ denotes variable parameters. In many cases, the summary statistic is formed through training a neural network on an event-level binary classification task to distinguish signal events from background events. The final summary statistic is thus the composition $T_{(\rho,\alpha)} = A_\alpha \circ R_\rho : x \mapsto \hat{z} \mapsto t$. As summaries are in general lossy, results inferred from them are usually weaker than those that would be obtained if the full likelihood were available. An important question in particle physics is thus the optimization of the summaries and in particular their parameters $(\rho,\alpha)$.

It is notable that both HEP and modern machine learning workflows based on foundation models exhibit a number of similarities regarding their use and optimization. The correspondence is sketched in Figure~\ref{fig:FM_HEP} and we describe it briefly in the next section.

\subsection{HEP in the Language of Foundation Models}
\label{sec:hep_foundation}

In modern ML practice based on foundation models, training often proceeds through two phases. In the first phase, models are trained on a large \emph{pretraining dataset} using \emph{pretext tasks}. These tasks often do not solve the task for which the model is ultimately used, but rather are designed in order to allow the model to create useful, semantically meaningful representations from low-level input data. Here, the models can be split into a \emph{backbone} model that forms the representations and a \emph{pretext head} that outputs the final prediction of this training stage.

In a second phase, the pretrained backbone (i.e. the model with the pretext head removed) is adjusted for the target downstream task by combining the backbone model with a suitable prediction head component and the resulting composite model is trained on the \emph{target dataset}. Here two training strategies can be pursued that differ in computational complexity. In one mode, the backbone acts as a fixed feature extractor and only the head is optimized for the new downstream task. Alternatively, the backbone weights can be included in the second-stage optimization to yield a feature extractor that is \emph{finetuned} to the downstream task at hand. Both of these strategies are to be contrasted to the ``from-scratch'' strategy, in which no pretraining occurs and the full composite model of backbone and target head is only optimized using the target dataset.

The optimization of a reconstruction and analysis pipeline for data analysis in HEP proceeds along very similar directions. Reconstruction can be interpreted as a backbone model designed to provide physicists, interested in downstream physics tasks, with a useful general-purpose representation of the low-level event data. Viewed through this lens, we can recognize the reconstruction algorithms as feature extractors that are optimized on pretext tasks, usually in a supervised manner where the reconstructed event record $\hat{z}$ is optimized to estimate the latent event record $z$.

\begin{equation}
\rho^* = \underset{\rho}{\mathrm{argmin}}\;\mathbb{E}_{p(x,z)} \mathcal{L}_\mathrm{rec.}(\hat{z} = R_\rho(x),z)    
\end{equation}

Such pretext tasks include predicting (i.e. reconstructing) e.g. kinematic variables of the particles within the latent states, the particle type, or the true flavor of jets. Similarly to the pretraining dataset in ML workflows, the optimization of these algorithms is often carried out using large simulated samples of particle collisions that may not be used in the final analysis.

\begin{table}
    \centering
    \begin{tabular}{p{0.2\textwidth}p{0.26\textwidth}}
        \toprule
         ML & HEP \\\midrule
         Foundation Model & Reconstruction \\
         Pretext Tasks & Reconstruction Closure \\
         Downstream Head & Analysis\\
         Finetuning & Analysis-specific Reconstruction choices,  Working Points\\
         Embedding & Object Observables \\
         \bottomrule
    \end{tabular}
        \caption{Shared concepts between modern Machine Learning with foundation models and current practice in High-Energy Physics}
    \label{tab:my_label}
\end{table}

The downstream task in HEP is the analysis stage in the HEP pipeline where the extracted features, i.e. the reconstructed event, are used as inputs to compute a suitable summary statistic. The setup is most similar to the ``frozen backbone'' model, where the event representation is fixed and only the downstream analysis itself is optimized for a physics task, such as a measurement of a particle property or a search for new particles. Here, the fixed reconstruction with parameters $\rho^*$ induces a distribution $p_{\rho^*}(\hat{z}, \theta)$ for which samples are available to optimize the analysis
\begin{equation}
{\alpha^*}_{|\rho^*} = \underset{\alpha}{\mathrm{argmin}}\;\mathbb{E}_{p_{\rho^*}(\hat{z},\theta)} \mathcal{L}_\mathrm{ana.}(t = A_\alpha(\hat{z}),\theta)    
\end{equation}
Here, it is important to note that the sequential (greedy) optimization strategy of first optimizing the reconstruction and then the analysis does not necessarily coincide with the joint optimum
\begin{align}
\begin{split}
    (\rho^*, {\alpha^*}_{|\rho^*}) \neq ({\rho^*}_\mathrm{joint},{\alpha^*}_\mathrm{joint}) = \\\underset{{\rho,\alpha}}{\mathrm{argmin}}\;\mathbb{E}_{p(x,\theta)} \mathcal{L}_\mathrm{ana.}(t = A_\alpha(R_\rho(x)),\theta)
\end{split}
\end{align}
Thus, a joint optimization e.g. through finetuning would in general be desirable.

While in general the reconstruction is thought of largely as a static summary, some sub-algorithms within it may be available in a discrete number of well-defined configurations referred to as ``working points''. It is common practice for analyzers to select a configuration particularly suited for their specific physics analysis among this discrete set of options. Viewed from a ML perspective, this may be interpreted as a basic approach to non-gradient-based finetuning.

Based on these correspondences, we can recognize an opportunity for a more complete and automated finetuning of reconstruction-level components,
in the context of a joint optimization of the full analysis pipeline. In light of the general trends towards larger neural network components~\cite{Pata:2022wam,ExaTrkX:2021abe,DiBello:2022iwf} and advances in differentiable programming, gradient information of the output of reconstruction algorithms with respect to their configuration parameters becomes increasingly accessible. Hence, a \emph{gradient-based} finetuning and \emph{joint optimization} as it is common in machine learning becomes possible by computing the gradient of the final event-level loss (e.g. binary signal vs. background classification) with respect to all differentiably connected components at both the analysis-level and reconstruction-level. 

In addition to the optimizations of the algorithms themselves, the choice of features that describe objects within the reconstruction is a regular target of optimization. For example, new jet-level observables, such as jet substructure variables~\cite{Thaler:2010tr, Marzani:2019hun} or jet-tagging scores may be added to the reconstruction output if such features aid the downstream analysis-level processing. This choice is not unlike the choice of embedding dimension of the backbone output within ML foundation models. In this context, it is interesting to explore to what extent learned, instead of hand-engineered features may aid downstream performance and how they can be finetuned.

\section{Demonstrator Model and Dataset}
\label{sec:testcase_and_data}

We demonstrate the concept of analysis-level finetuning of neural reconstruction components in a simplified setting of a new resonance, graviton $G$, decaying to two Higgs Bosons, which in turn decay through the $H\to b\bar{b}$ channel. The final state to be analyzed is thus a multi-jet final state with $G\to HH \to b\bar{b}b\bar{b}$.
A typical analysis strategy would be split into two stages. At the reconstruction-level, a ``$Xbb$ tagging algorithm'' would typically be developed, i.e. a binary classifier that operates on the constituents of a large-radius jet and infers whether it originated from a $H\to b\bar{b}$ decay. At the analysis-level, jets within the event would be analyzed to perform a full event classification to determine whether the event originated from a signal or background process such as multi-jet backgrounds. The work primarily investigates to what extent the reconstruction-level jet processing can be \emph{finetuned} to yield an improved full-event classification performance. For the study two datasets are primarily used, which we describe briefly:

\begin{itemize}
    \item[] \textbf{JetClass} This dataset~\cite{jetclass} consists of 100M simulated anti-$k_\text{T}$ R=0.8 \cite{Cacciari:2008gp} ``large-R'' jets initiated from 10 different decay configurations of heavy states, including $H\to b\bar{b}$. This dataset is only implicitly used through the reuse of the published pretrained network weights in the domain adaptation studies. The decays were simulated through \texttt{MadGraph}~\cite{MadGraph} and the parton shower evolution of the final-state particle was simulated via \texttt{Pythia}~\cite{Pythia}. The final data was then prepared through the \texttt{Delphes}~\cite{deFavereau:2013fsa} simulator. 

    \item[] \textbf{CMS Open Data} This dataset~\cite{open-data} consists of 10M simulated events divided into QCD background and $G\to HH$ signal, where the signal is a mixture of X mass points from 600 GeV to 4500 GeV. For pre-training on the $Xbb$ jet task we use the dataset as a jet dataset for a total of 22M jets, while for end-to-end training we reshape the data such that data instances are full events with multiple jets. As  the provided CMS dataset saves the jet-level information, we edited the \texttt{HiggsToBBNtupleProducerTool}~\footnote{The repository released with the CMS $Xbb$ tagging dataset: \url{https://github.com/cms-opendata-analyses/HiggsToBBNtupleProducerTool}} released with the dataset to also save this event-level information. 
    We consider a loose event selection criteria, keeping events with at least two large-R jets with $p_\text{T} > 150$~GeV. For the analysis classifier, we consider the five highest $p_\text{T}$ jets in the event, which keeps 99.5\% of the true $H\to b\bar{b}$ jets in these graviton signal samples. When reporting performance on the analysis classifiers, these cuts define the denominator of the signal and background efficiencies. The dataset has been produced through the simulation and reconstruction pipeline of the CMS experiment~\cite{Chen:2021euv}.
\end{itemize}

\section{Architecture and Training Strategies} 
\label{sec:model}

We analyze the setup described above along two dimensions: architectural constraints and training strategies, with the goal to explore how much performance in downstream tasks can be gained by moving beyond the traditional HEP workflow.

\subsection{Architectures}

\begin{figure}
    \centering
    \includegraphics[width=0.45\textwidth]{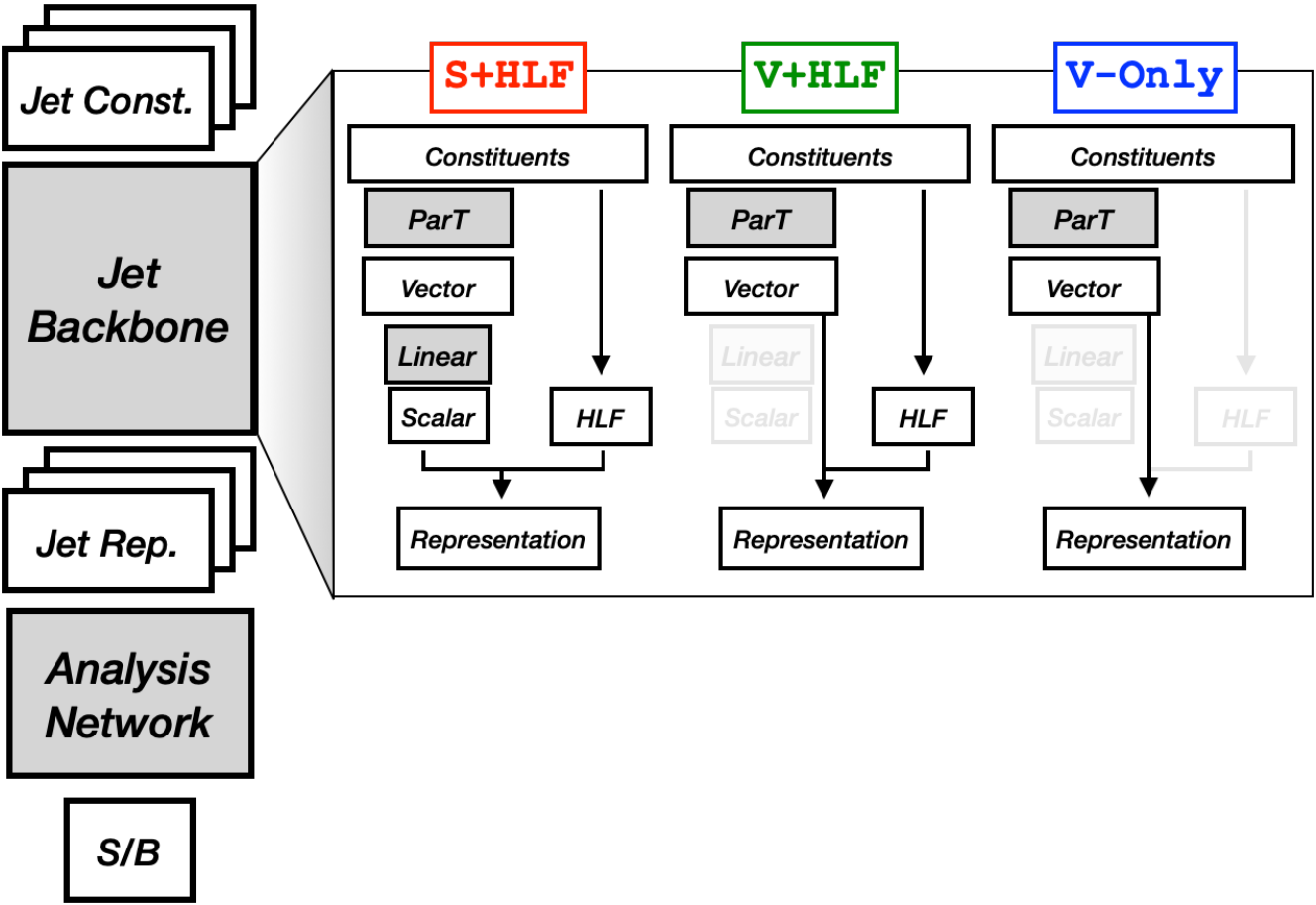} % Figures/Architectures.pdf
    \caption{Hierarchical neural network structures considered in this work with decreasing levels of structural constraints and manually engineered features.}
    \label{fig:Architectures}
\end{figure}

Overall we investigate three possible architectures with a decreasing amount of interpretable physics structure to determine how much structure and manually engineered features are needed, or whether generic high-dimensional learned representations would suffice.

The networks consist of a reconstruction-level network for jets that operates on constituents, and optionally may be augmented with physics-driven high-level features (HLF), to construct a jet representation. These representations of all jets within the event then enter a permutation-invariant analysis-level network. For the reconstruction-level network we use the transformer-based \texttt{ParT} architecture (sans the final softmax layer) to form embeddings of the jet constituents, which may optionally be projected to a scalar value through a linear layer. With the \texttt{ParT} network, we use the same inputs as proposed in Ref~\cite{Qu:2022mxj}.
For the analysis-level a deep set~\cite{2017arXiv170306114Z, Komiske:2018cqr} architecture is used to reflect lack of inherent ordering to the reconstructed jets. The choice is made for simplicity and additional performance may be achieved through more complex permutation-invariant architectures such as transformer networks. The architectures differ in the details on how the jet representation is formed as shown in Figure~\ref{fig:Architectures}, each progressively removing physics-motivated features and data flow in lieu of a less structured architecture.

\begin{itemize}
    \item[] \textbf{Jet Scalar + HLF (\texttt{S+HLF})}: This is the traditional HEP architecture, where particle jets are described by a small number of high-level features (HLF), which we keep to a minimum with five features: the kinematic variables $p_\mathrm{T}$, $\eta$, $\phi$, the jet mass $m_j$ and the soft-drop mass $m_\mathrm{sd}$ \cite{Larkoski_2014}. In addition to these fixed features, we add a slot for one additional \emph{learned} scalar feature. From the physics at hand, we could motivate to use this scalar value to e.g. add the classifier output of a (a pretrained) $Xbb$-tagging algorithm. In the \texttt{finetuned} and \texttt{from-scratch} training configurations described below, the network may choose to use this learnable slot to propagate other summaries of the jet constituent through this bottleneck that may not correspond to a $Xbb$-score.

    \item[] \textbf{Jet Vector + HLF (\texttt{V+HLF})}: Instead of only allocating a single scalar, here the analysis-level network can circumvent the scalar bottleneck and access the raw latent vector representation of the reconstruction-network without the final projection to a scalar value. This may enable the analysis-level network to make use of a richer representation of the jet.

    \item[] \textbf{Jet Vector (\texttt{V-Only})}: When the analysis-level network has access to a high-dimensional embedding of the constituents, one may hypothesize that the high-level jet features may not be needed, as the corresponding information is already encoded within the latent embedding of the \texttt{ParT} backbone. In this architecture we drop all HLF and just use the latent jet embedding to pass information to the analysis-level network.
\end{itemize}

\subsection{Training Strategies}

We pair the three architectures with three training strategies for the combined network consisting of reconstruction- and analysis-level components. The overall goal of the composition is to optimize on binary classification of the signal process against multijet background as measured through standard binary cross-entropy.

\begin{itemize}
    \item[] \textbf{Frozen Pretraining (\texttt{frozen})}: This model resembles the traditional HEP workflow. The jet backbone model is trained on a reconstruction-level task and then frozen. The pretext task for this pretraining is the classification of jets as originating from a $X\to b\bar{b}$ decay chain and the model is randomly initialized:
    The pretrained jet backbone is then integrated into the analysis as a frozen feature extractor and only the analysis-level network is optimized on the resulting jet representation. In the \texttt{S+HLF} the additional learned slot is then populated with the classifier output, whereas in the \texttt{V+HLF}  and \texttt{V-Only} the latent representation of the $Xbb$-tagger just before the classification head passed to the analysis.
    
    \item[] \textbf{Finetuned Training (\texttt{finetuned})}: In this model, the backbone is initialized to the pretrained weights, but during the training, gradient information is propagated to both the analysis- and reconstruction-level networks. That is, the jet-backbone is allowed to adapt to the specific analysis environment to minimize the analysis-level loss. Thus, while e.g. in the \texttt{S+HLF} model, at initialization time, the scalar value passed to the analysis is exactly the $Xbb$ score, during training the semantic meaning of this neuron may drift as the network learns to encode other types of information as well, making use of the notion of polysemanticity in neural networks~\cite{2023arXiv230409707O}.

    \item[] \textbf{No Pretraining (\texttt{from-scratch})}: To assess the impact of pretraining and finetuning we train the full composed network end-to-end from randomly initialized weights only on the final analysis-level classification task. In this model, the network is completely free to choose what information to propagate through the latent states. In particular, the scalar value in the \texttt{S-HLF} model need not be related to the probability of originating from a $Xbb$-decay.
\end{itemize}

\noindent In order to avoid vanishing gradients impeding efficient gradient-based training we remove the sigmoid activation in the \texttt{S+HLF} models for the \texttt{finetuned} and \texttt{from-scratch} configurations. The \texttt{ParT} backbone is trained following the training setup of the original paper: a binary cross-entropy loss is minimized by means of the Lookahead optimizer \cite{zhang2019lookahead} with $k = 6$ and $\alpha = 0.5$, and RAdam as inner optimizer \cite{liu2021variance} with $\beta_1 = 0.95$, $\beta_2 = 0.999$, and $\epsilon = 10^{-5}$. The same setup is adopted when training the full pipeline end-to-end together with the analysis head network, while when training the head alone on a frozen jet representation we use the Adam optimizer~\cite{2014arXiv1412.6980K}. A batch size of 512 for the backbone pretraining, 256 for the end-to-end analysis model, and a starting learning rate of 0.001 is employed, using a constant learning rate scheduler with warm-up whenever the backbone parameters are learnable. A model checkpoint is saved after every epoch and the one with lowest loss on the validation set is chosen for the final performance evaluation on the test set. The datasets are divided into training, validation, test dataset with a 45\% / 5\% / 50\% split.

\section{Results}
\label{sec:results}

\begin{figure}[h]
    \centering
    \includegraphics[width=0.37\textwidth]{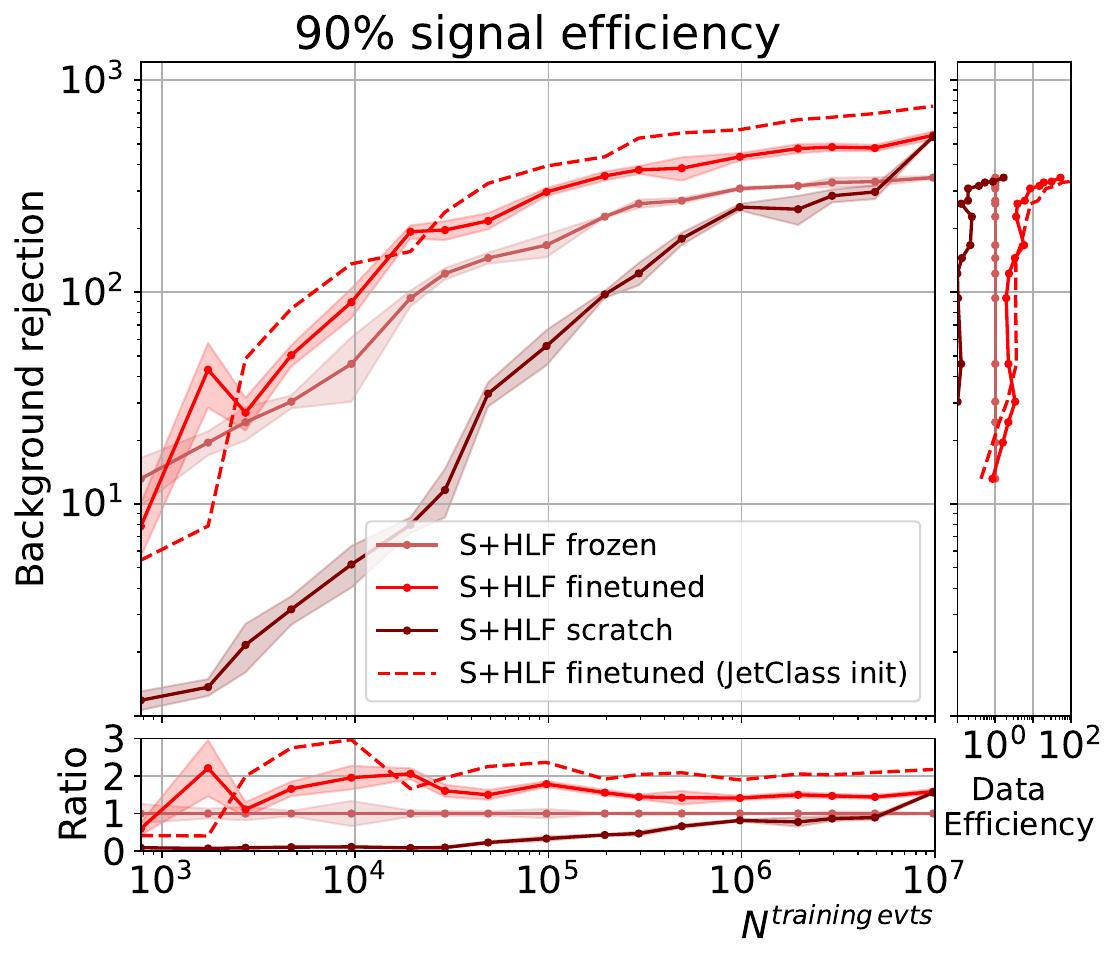}
    \includegraphics[width=0.37\textwidth]{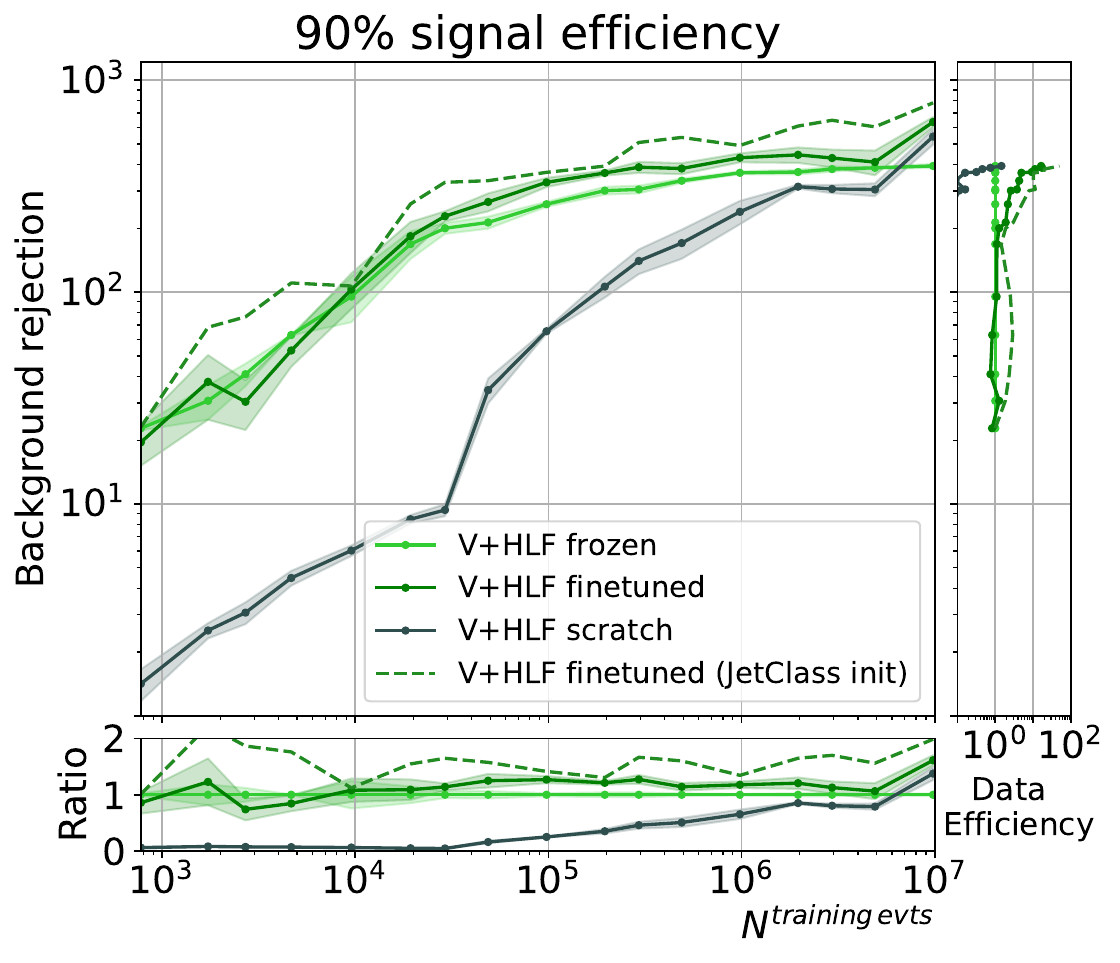}
    \includegraphics[width=0.37\textwidth]{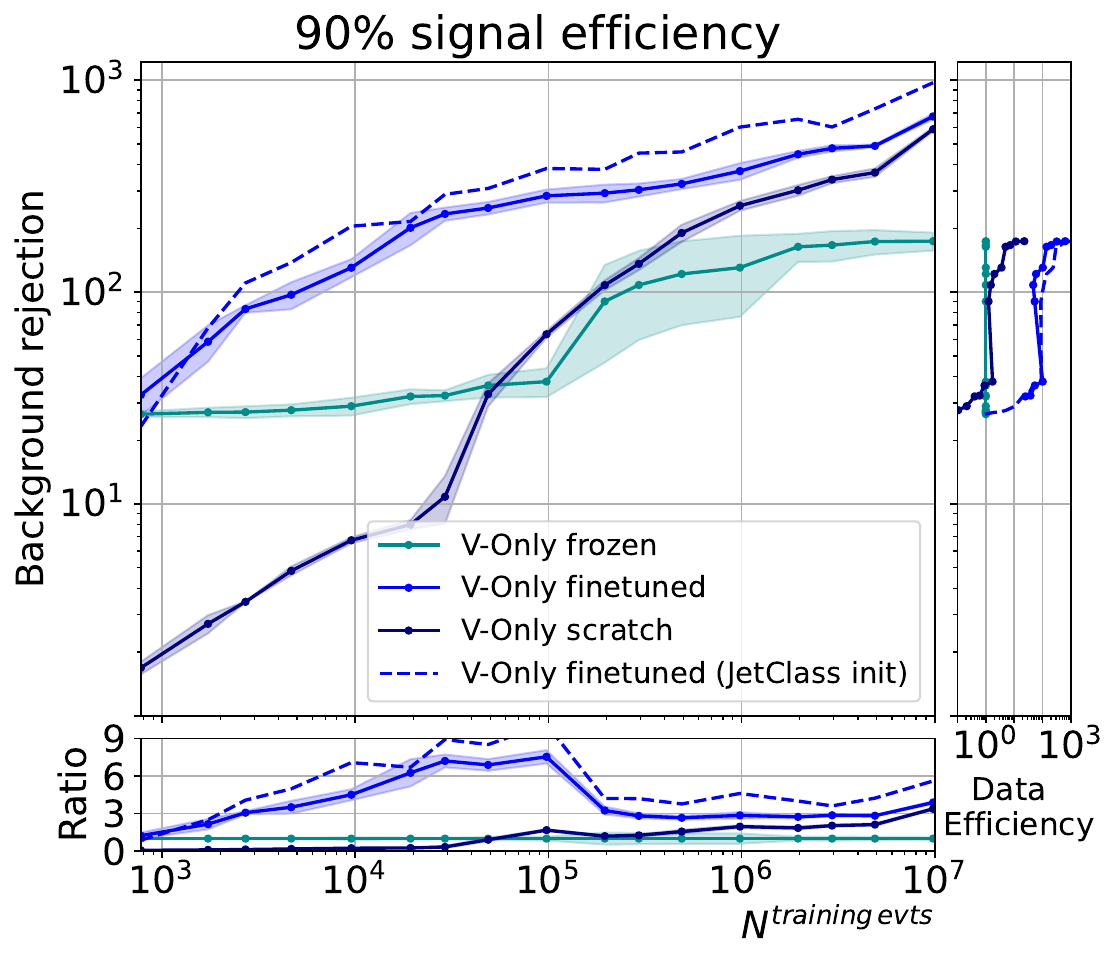}
    \caption{Performance as a function of labeled examples across three training strategies shown for the investigated architectures. For all architectures we see a significant benefit from finetuning over a frozen backbone. Pretraining is significantly more performant than training from scratch. For very large datasets from-scratch training can exceed a frozen backbone.}
    \label{fig:finetuning_is_good}
\end{figure}

We present the results primarily through comparing performance as a function of labeled examples in the final analysis-level signal-vs.-background (S/B) classification task. Reported signal efficiencies are inclusive over all graviton resonance masses. The shown uncertainty bands are based on the standard deviation of four independent runs. As a baseline model, we will then compare the results of the various combinations of architecture and training regimes to the \texttt{S-HLF(frozen)} setup, as this resembles the standard HEP workflow of a fixed reconstruction on top of which an analysis is optimized most closely. 
% The main performance metric here is chosen to be the background rejection (i.e. the inverse false positive rate). 
The main performance metric presented here is the background rejection (i.e. the inverse false positive rate).
Results on alternative measures are presented in the appendix. An increased level of performance can be interpreted in two dimensions:

\begin{itemize}
    \item[]\textbf{Fixed Dataset Size}: Here we compare the performance of the trained model at a fixed number of training examples for the downstream analysis-level task.
    
    \item[]\textbf{Fixed Performance Level}: Alternatively, it is interesting to explore the number of training samples required for a given performance level. Two models may be able to reach the same performance but the more data-efficient one will require less training examples and thus computational resources to reach it. We define the data efficiency as the ratio of the required dataset size to reach a performance level as compared to that of the baseline.
\end{itemize}

\subsection{Training Strategy Comparison}

\begin{table}
    \centering
    \begin{tabular}{rccc}
    \toprule
      & \texttt{S+HLF}  & \texttt{V+HLF} & \texttt{V-Only} \\
      \midrule
      % & \textcolor{red}{\texttt{S+HLF}}  & \textcolor{Green}{\texttt{V+HLF}} & \textcolor{blue}{\texttt{V-Only}} \\ \hline
     \texttt{frozen}    & 350$\pm$10  & 390$\pm$10  & 170$\pm$20 \\ 
     \texttt{finetuned}    & 550$\pm$20 & 640$\pm$40 & \textbf{680$\pm$20} \\ 
     \texttt{from-scratch}   & 540$\pm$10 & 540$\pm$50 & 590$\pm$10 \\
     \bottomrule
    \end{tabular}
    \caption{Background rejection at 90\% Signal Efficiency for the nine investigated configurations.}
    \label{tab:perf_at_fixed_N}
\end{table}

\begin{table}
    \centering
    \begin{tabular}{rccc}
    \toprule
      & \texttt{S+HLF}  & \texttt{V+HLF} & \texttt{V-Only} \\ \hline
     \texttt{frozen}    & 1  & 14.00  & / \\ 
     \texttt{finetuned}    & 53.00 & \textbf{67.00} & 14.00 \\ 
     \texttt{from-scratch}   & 1.70 & 1.70 & 2.80 \\
     \bottomrule
    \end{tabular}
    \caption{Data efficiency with respect to \texttt{S+HLF} frozen model at 90\% Signal Efficiency for the nine investigated configurations. }
    \label{tab:dataeff}
\end{table}

In Figure~\ref{fig:finetuning_is_good}, we first compare the performance of each of the different architectures under the suite of training strategies described above. Here, we expect the pretraining to clearly outperform from-scratch training as the pretext task is strongly suggested by the physics at hand. The relationship between finetuned and frozen backbones, however, is less clear. While the frozen backbone should provide a lower bound on the finetuning performance, the level of performance gain that finetuning may achieve depends strongly on the alignment of the pretext task and its learned representations with the downstream task. For example, if the pretrained representation of the jets within the event would be a sufficient statistic on the inference target, finetuning would not be able to extract any more information from the low-level data. In the present example, however, we do observe a significant gain from finetuning, which manifests in a increase in background rejection e.g. a from 1.5-4x at 90\% signal efficiency as shown in Table~\ref{tab:perf_at_fixed_N}. Expressed in terms of data efficiency, the finetuned models reach a high level of performance with up to 70x less data as shown in Table~\ref{tab:dataeff}. Training the full architectures from scratch reaches high levels of performance but requires significantly more labeled examples. We point out that from-scratch training, when trained on sufficient data, eventually surpasses the performance of frozen backbone models, further indicating that the frozen jet-level representations may not be sufficient.

\begin{figure}
    \centering
    \includegraphics[width=0.45\textwidth]{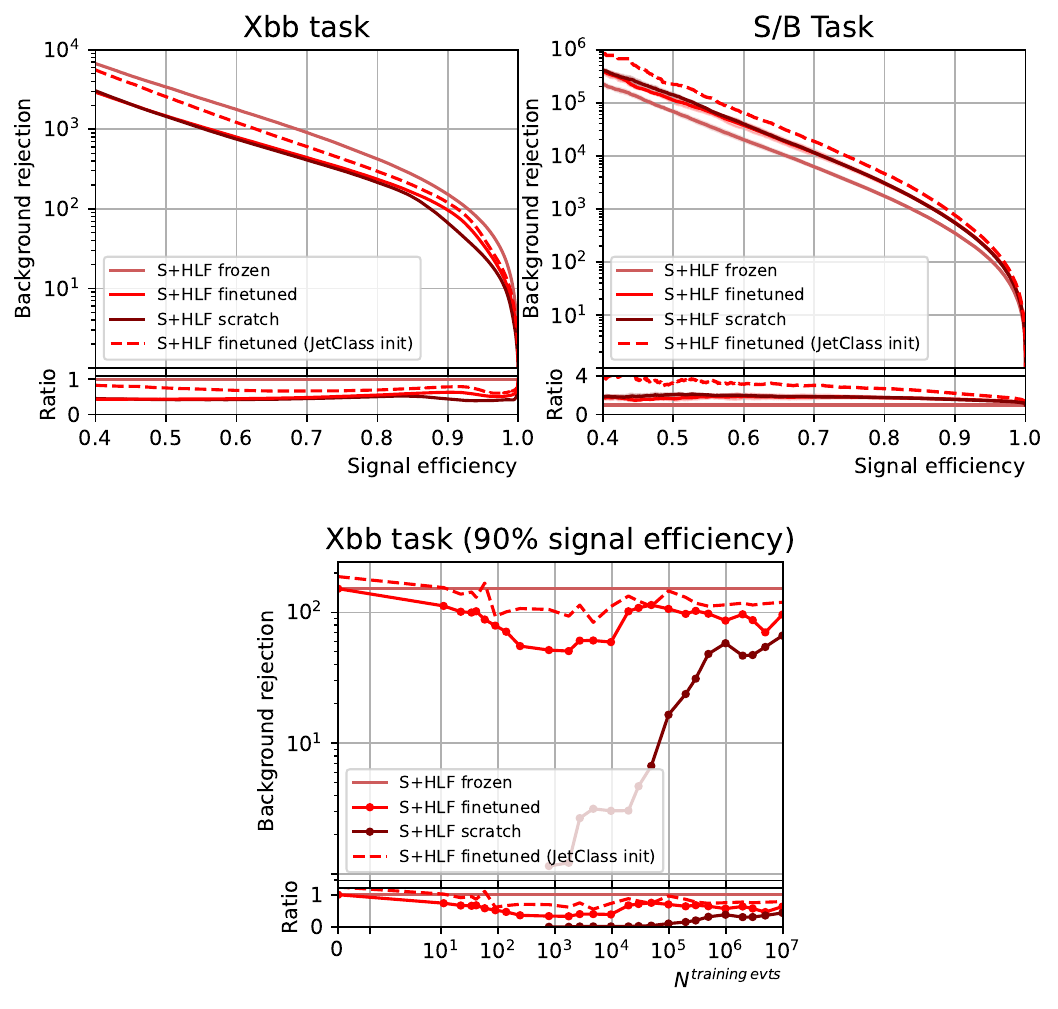}
    \caption{Top: Performance metrics of \texttt{S+HLF} for pretext (left) and downstream (right) tasks. In \texttt{finetuned} training the learnable scalar in \texttt{S+HLF} trades off $Xbb$ performance against downstream task performance. In \texttt{from-scratch} training $Xbb$-tagging emerges as a useful subtask without supervision. Bottom: $Xbb$ Performance of learned scalar feature as function of training samples}
    \label{fig:compare_concept_drift}
\end{figure}

We also explore the drift of the learned feature in the \texttt{S+HLF} models. In the frozen backbone, this scalar represents the probability of the jet to originate from a $H\to bb$ decay. In the \texttt{finetuned} and \texttt{from-scratch} configurations this interpretation may not hold anymore, as the continued training of the jet-level backbone may overload this neuron semantically. We can investigate this learned scalar through the lens of an $X\to bb$ classifier by adding a sigmoid activation to the scalar output of the non-frozen \texttt{S+HLF} models. We observe that indeed during finetuning the learned scalar feature drifted and its $Xbb$ performance deteriorated, while the overall performance of the finetuned models surpasses the frozen model as shown in Figure~\ref{fig:compare_concept_drift}. Hence we hypothesize that during learning, the learned scalar is overloaded to encode multiple jet features relevant for the downstream task.

\begin{table}
    \centering
    \begin{tabular}{crr}
        \toprule
        & Bkg. rej. & Ratio \\ 
        \midrule
        \texttt{frozen} & 151.93 & 100.00\% \\
        \texttt{finetuned} & 96.28 & 63.38\%\\
        \texttt{from-scratch} & 66.53 & 43.79\% \\
        \texttt{finetuned} (\texttt{JetClass} init) & 119.67 & 78.77\% \\
        \bottomrule
    \end{tabular}
    \caption{Performance of the scalar feature at 90\% signal efficiency in trained \texttt{S+HLF} networks on $Xbb$-tagging}
    \label{tab:drift}
\end{table}

It is interesting to note that the learned scalar from the \texttt{S+HLF(from-scratch)} model performs non-trivially at the $Xbb$ tagging task, without ever having received a feedback from ground-truth labels of the jets. That is the importance of the $Xbb$ sub-task \emph{emerges autonomously} in the end-to-end learned models after around $10^4-10^5$ training examples as shown in the bottom pane of Figure~\ref{fig:compare_concept_drift}. The end-to-end model ultimately achieves a performance of 43\%  of the supervised $Xbb$-pretraining as shown in Table~\ref{tab:drift}.

\subsection{Architecture Comparison}

\begin{figure}
    \centering
    \includegraphics[width=0.42\textwidth]{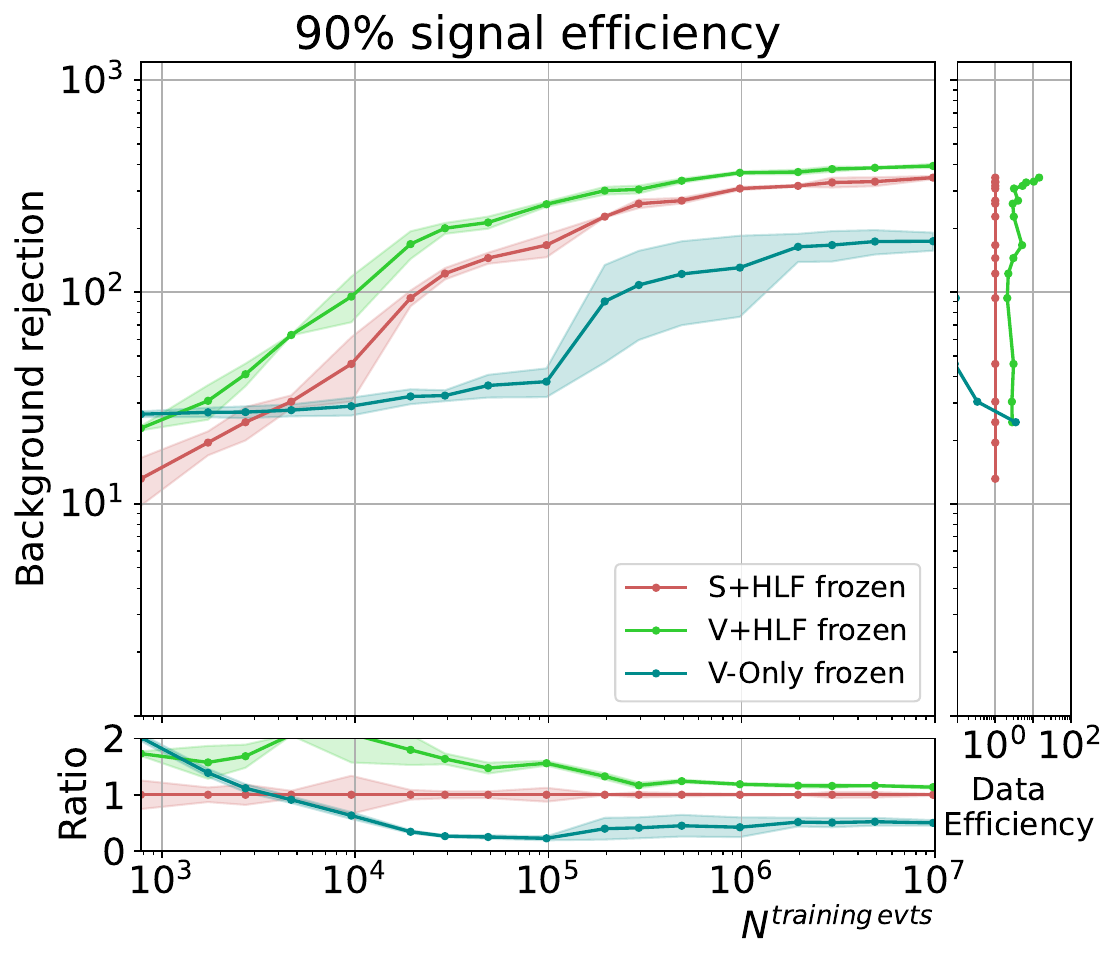}
    \caption{Performance metrics for \texttt{frozen} configuration across architectures. We observe that higher-dimensional embeddings show improved performance.}
    \label{fig:vector_is_better}
\end{figure}

We now compare the performance of the different architectures under a fixed training strategy to assess to what extent the models 
% with less physics-informed architecture 
with less physics-information % more concise reads better to my ear
can learn representations that are more effective at the downstream task. The results for the frozen training strategy are shown in Figure~\ref{fig:vector_is_better}, indicating that with a fixed backbone the higher-dimensional embeddings do indeed carry more information than just the scalar Xbb score. However, they seem to not fully capture the information contained within the high-level features. This result renders the \texttt{V+HLF(frozen)} model the best performing with an improved background rejection at 90\% signal efficiency that is 14\% higher than \texttt{S+HLF(frozen)}. Furthermore, the model is up to 15$\times$ more data efficient than the baseline model. While the \texttt{Vector-Only(frozen)} model initially outperforms the baseline, with sufficient training data, the baseline model eventually surpasses it in performance. For the \texttt{finetuned} and \texttt{from-scratch} trained models, where the latent representation of the backbone can be adjusted to the downstream task, the missing information can be recovered, as shown in Figure~{\ref{fig:finetune_highd_is_less_important}} in the appendix. Hence, both the \texttt{Vector-Only} and \texttt{S+HLF} models generally reach the same level of performance with only minimal difference to the \texttt{Vector+HLF} models.

\subsection{Domain Adaptation}

\begin{figure}
    \centering
    \includegraphics[width=0.42\textwidth]{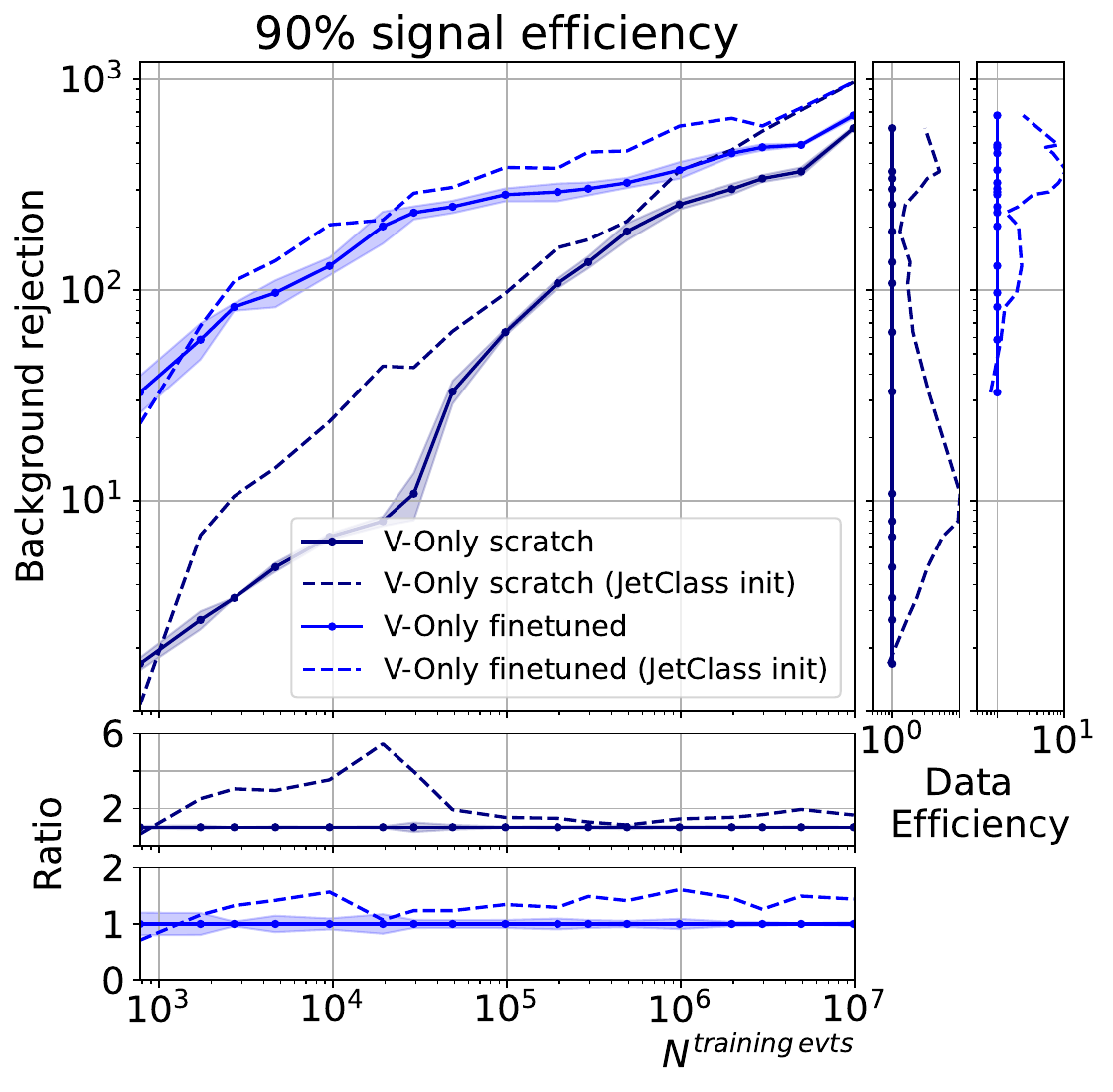}
    \caption{Initializing the jet-level networks in the Xbb pretraining (\texttt{finetuned}) or the end-to-end downstream task training (\texttt{from-scratch}) with the \texttt{JetClass}-trained network parameters boosts performance significantly.}
    \label{fig:double_finetune_it}
\end{figure}

\begin{table}
    \centering
    \begin{tabular}{rcc}
    \toprule
      & \texttt{random init}  & \texttt{JetClass init} \\
      \midrule
     &\multicolumn{2}{c}{Performance at $N_\mathrm{train}$=10M } \\
      % & \textcolor{red}{\texttt{S+HLF}}  & \textcolor{Green}{\texttt{V+HLF}} & \textcolor{blue}{\texttt{V-Only}} \\ \hline
     \texttt{V-Only (from-scratch)}    & 590$\pm$10  & 970 \\ 
     \texttt{V-Only (finetuned)}    & 680$\pm$20 & 970 \\
     \midrule
     &\multicolumn{2}{c}{Data Efficiency} \\
     \texttt{V-Only (from-scratch)}    & 1 & 1.70 \\ 
     \texttt{V-Only (finetuned)}    & 1 & 1.40 \\
     \bottomrule
    \end{tabular}
    \caption{Background rejection and Data Efficiency at 90\% Signal Efficiency under random or \texttt{JetClass} initializtion of the pretraining (\texttt{finetuned)} or end-to-end trainind (\texttt{from-scratch)}. Data efficiencies are computed with respect to the random initialization result of a given architecture}
    \label{tab:domainadapt}
\end{table}

A key aspect of foundation models in modern ML practice is their ability to form representations that may be \emph{transferable} to new datasets, and similar notions are also relevant in HEP. For example, while the target dataset in the above study is only of moderate size (22M jets), there are much bigger datasets available for a similar task such as the \texttt{JetClass} dataset described in Section~\ref{sec:testcase_and_data}, which contains 100M Jets. While those datasets are generated using different simulators and thus do not match directly at the distribution-level, i.e. they represent different domains, the underlying physics is largely similar. Therefore, domain adaptation may be possible such that pretraining on datasets other than the target dataset benefits the overall performance. The parameters of a \texttt{ParT} network, optimized for a 10-way multi-class inference of the originating decay chain of the jets in the \texttt{JetClass} dataset, have been made publicly available together with the dataset release~\cite{Qu:2022mxj}. We can therefore add one additional variant to each of the three training strategies.

\begin{itemize}
    \item[] \textbf{\texttt{JetClass}-pretrained Initialization} (\texttt{JetClass init}): For the two strategies with pretraining on the $Xbb$ task on the \texttt{CMS Open Data} dataset, (\texttt{frozen} and \texttt{finetuned}), the pretraining itself is initialized not randomly but from the published weights resulting from the multiclass training on \texttt{JetClass}. Similarly, in the \texttt{from-scratch} case, where no pretraning happens on the target datasets, the end-to-end training is initialized with the published weights as well. 
\end{itemize}

\noindent As shown in  Figure~\ref{fig:double_finetune_it} and Table~\ref{tab:domainadapt} we observe a significant improvement in performance for the models initialized from \texttt{JetClass}-pretrained weights. The performance gain is present in both \texttt{finetuned} and \texttt{from-scratch} models. We note that successful domain adaptation may open up interesting opportunities to cross-experiment pretrained foundation models in particle physics. The \texttt{JetClass}-initialized finetuning configurations are also shown as dashed curves in Figure~\ref{fig:finetuning_is_good}, where this configuration is consistently the best performing one.

\section{Conclusions}
\label{sec:conclusions}

In this work we investigated the possibility of adapting large-scale machine learning workflows from foundation models to particle physics. To this end we first developed a conceptual connection between ideas from modern machine learning such as foundation models, pretraining, finetuning, pretext tasks and vector embeddings and those that are common during the optimization of a particle physics analysis, such as reconstruction, tagging and analysis. We then explore these ideas in a case study of a Beyond Standard Model search, where the signal is defined as a heavy resonance decaying to two Higgs bosons, which in turn each decay via $H\to b\bar{b}$. In particular, we focus on establishing a performance hierarchy between training strategies: to what extent is finetuning advantageous over a frozen backbone trained on physics-defined pretext task (here: $Xbb$-tagging) and how much does the physics-based pretraining help over a direct end-to-end training of the downstream task?

We observe that finetuning does indeed add significant performance to the models measured both at fixed dataset sizes as well as in data-efficiency. Depending on the finetuned models, the gain in rejection can be as much as a factor of two larger than the frozen backbone, and 10-100 times more efficient at achieving a desired level of performance.
At the same time, the gap from the \texttt{frozen} to \texttt{from-scratch} models is significant in both dimensions, but reduced with sufficiently many training examples, where models trained from scratch can surpass frozen models due to being able to adjust the reconstruction-level representation of low-level data.

We identify two important research questions that go beyond the scope of this work, but build on its result. First, in light of the apparent benefits of reconstruction-level finetuning with respect to a downstream analysis-level task, the question of integrating and automating calibration techniques becomes important. One of the major benefits of a common, frozen backbone is the ability to correct simulation towards calibration data, which would have to now be done in-situ. Second, we recognize the interplay between designing valuable pretraining task and the need for finetuning. Observing significant benefits from finetuning may suggest it would be possible to re-capture parts of the additional performance, by understanding their physical origin and designing better pretrained representation that go beyond e.g. simple $Xbb$-tagging. If successful, the gap between frozen and finetuned models may be closed. We leave both research questions to future work.

\section{Acknowledgements}

The authors thank Michael Kagan, Sam Klein and Francesco Di Bello for valuable discussions and a careful read of the manuscript. LH and NH are supported by the Excellence Cluster ORIGINS, which is funded by the Deutsche Forschungsgemeinschaft (DFG, German Research Foundation) under Germany’s Excellence Strategy - EXC-2094-390783311.


\label{refer}
\bibliography{./references.bib}

\clearpage
\appendix

\section*{Appendix}
\label{sec:appendix}
\begin{figure}[H]
    \centering
    \includegraphics[width=0.42\textwidth]{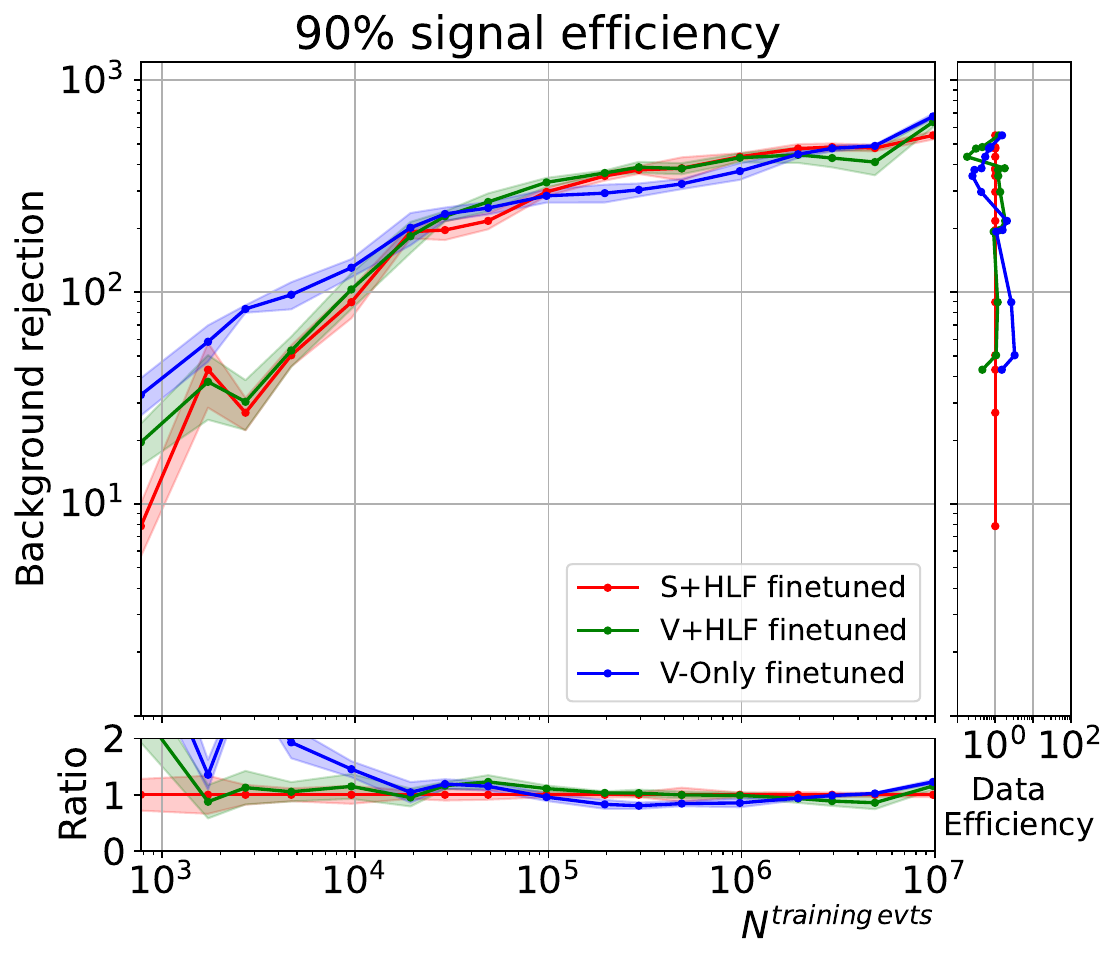}
    \includegraphics[width=0.42\textwidth]{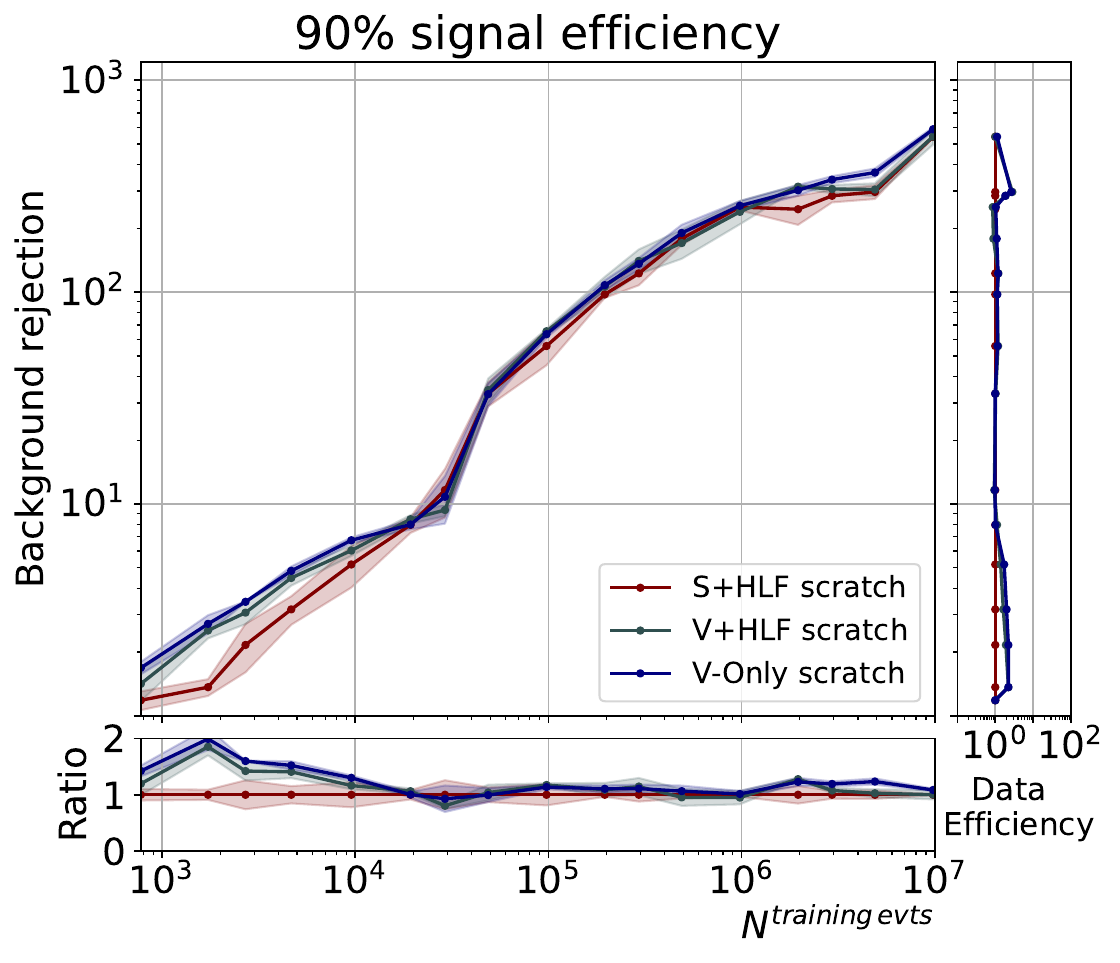}
    \caption{ %For frozen backbones, we win through access to latents...
    For the end-to-end optimized finetuned and from-scratch models, a single scalar encodes all of the relevant jet-level information.
    }
    \label{fig:finetune_highd_is_less_important}
\end{figure}

For the finetuned and from-scratch trained models, where the latent representation of the backbone can be adjusted to the downstream task, the missing information in \texttt{Vector-Only} and \texttt{S+HLF} frozen jet representations can be recovered, as shown in Figure~\ref{fig:finetune_highd_is_less_important}, reaching the same level of performance with only minimal difference to the \texttt{Vector+HLF} models.

\begin{figure}[H]
    \centering
    \includegraphics[width=0.42\textwidth]{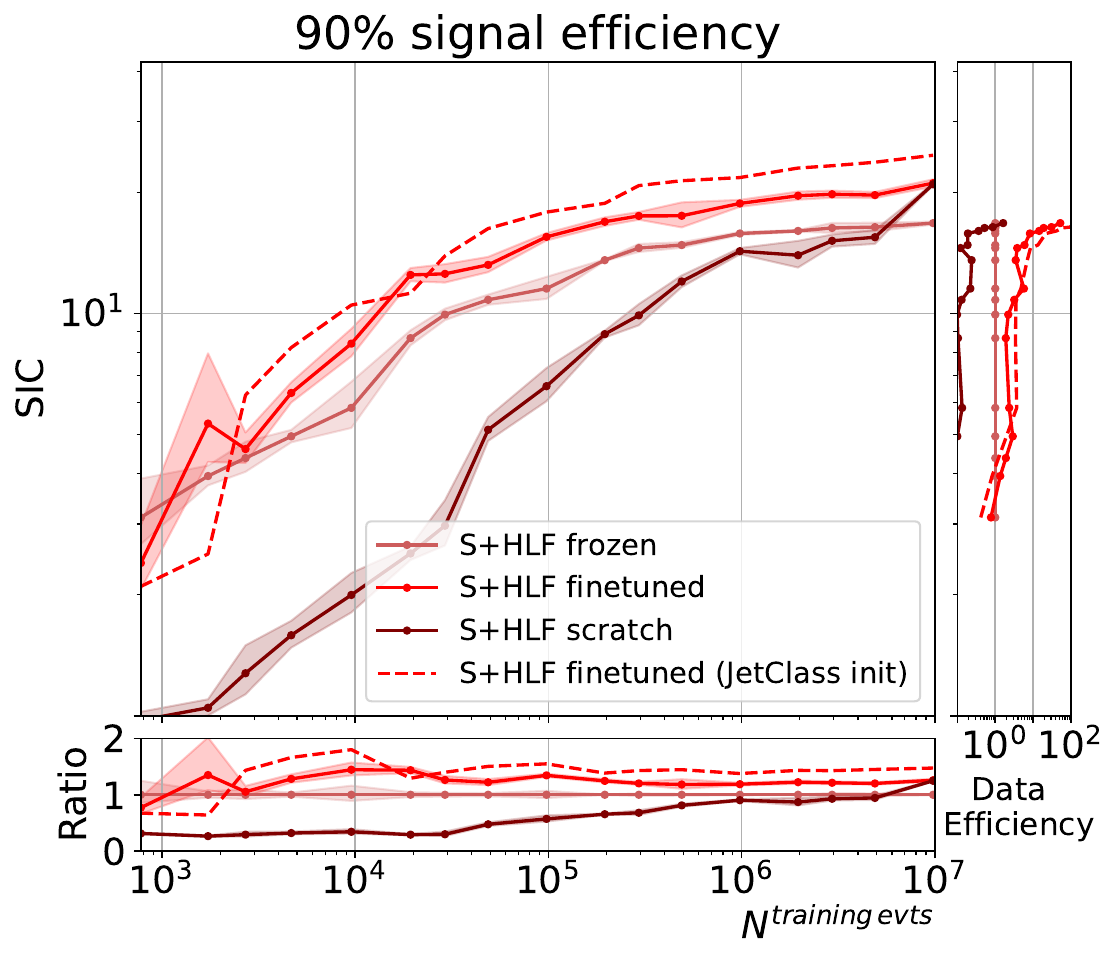}
    \includegraphics[width=0.42\textwidth]{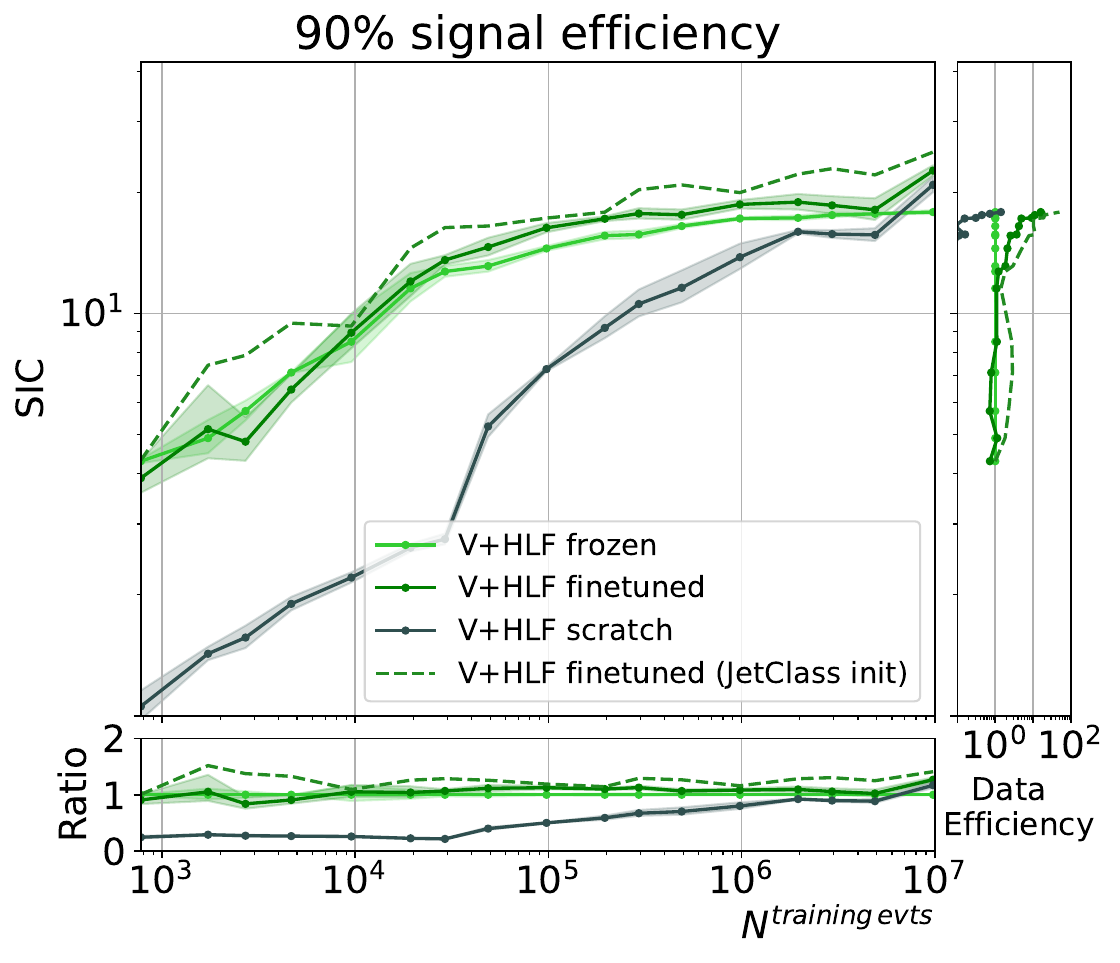}
    \includegraphics[width=0.42\textwidth]{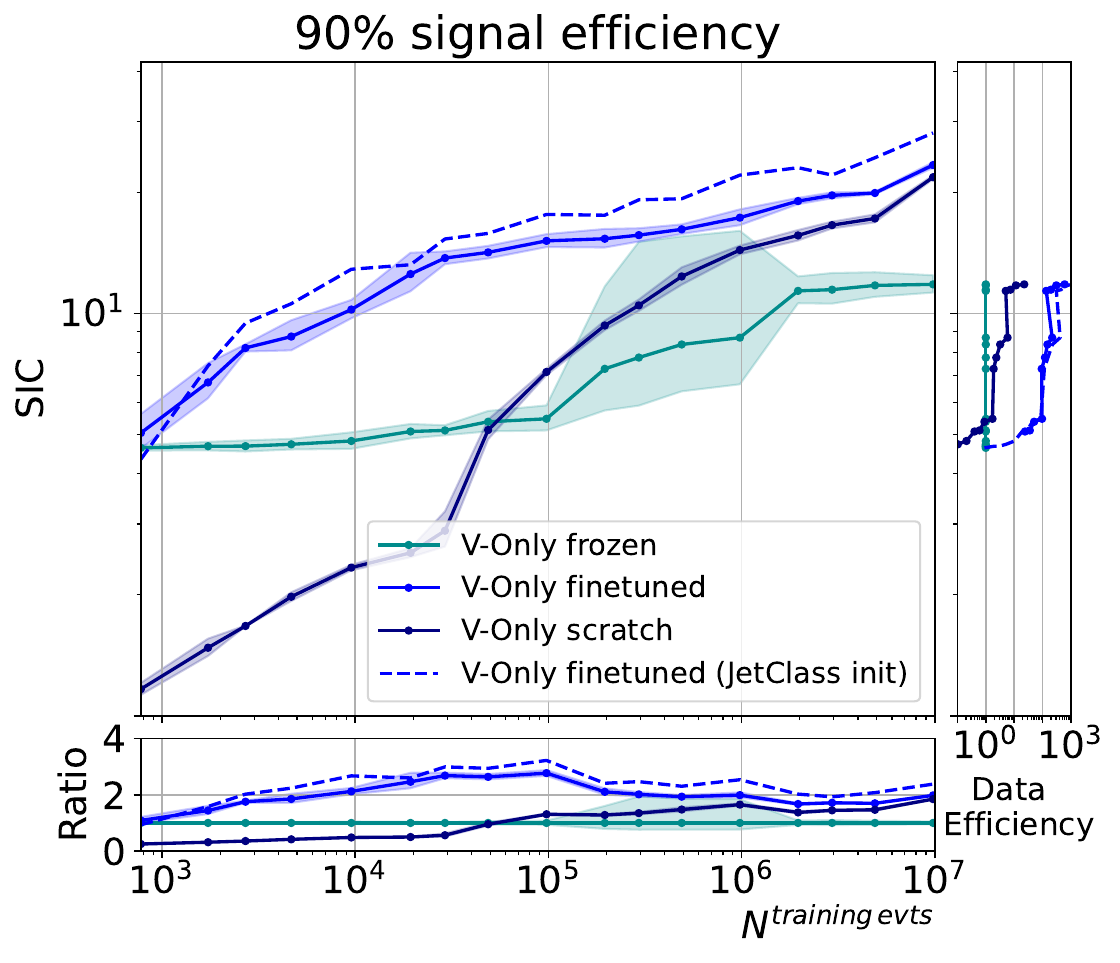}
    \caption{SIC performance as a function of labeled examples across three training strategies shown for the investigated architectures. For all architectures we see a significant benefit from finetuning over a frozen backbone. Pretraining is significantly more performant than training from scratch. For very large datasets from-scratch training can exceed a frozen backbone.}
    \label{fig:finetuning_is_good_SIC}
\end{figure}

\begin{figure}[H]
    \centering
    \includegraphics[width=0.42\textwidth]{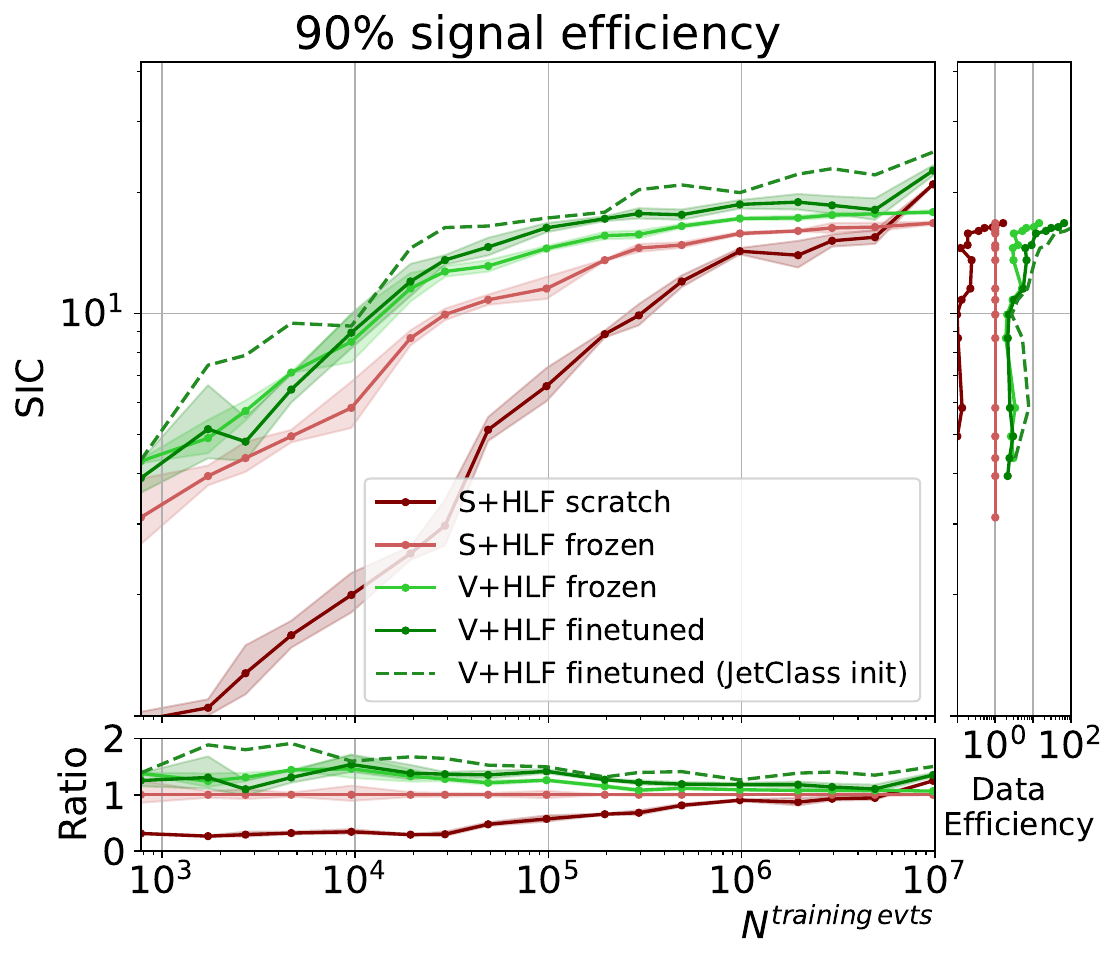}
    \caption{SIC performance as a function of labeled downstream examples. Methods from foundation models such as large-scale pretraining, finetuning, high-dimensional embedding yield significant benefits in performance and data efficiency over the baseline (\texttt{S+HLF}).}
    \label{fig:SIC}
\end{figure}

\begin{figure}[H]
    \centering
    \includegraphics[width=0.42\textwidth]{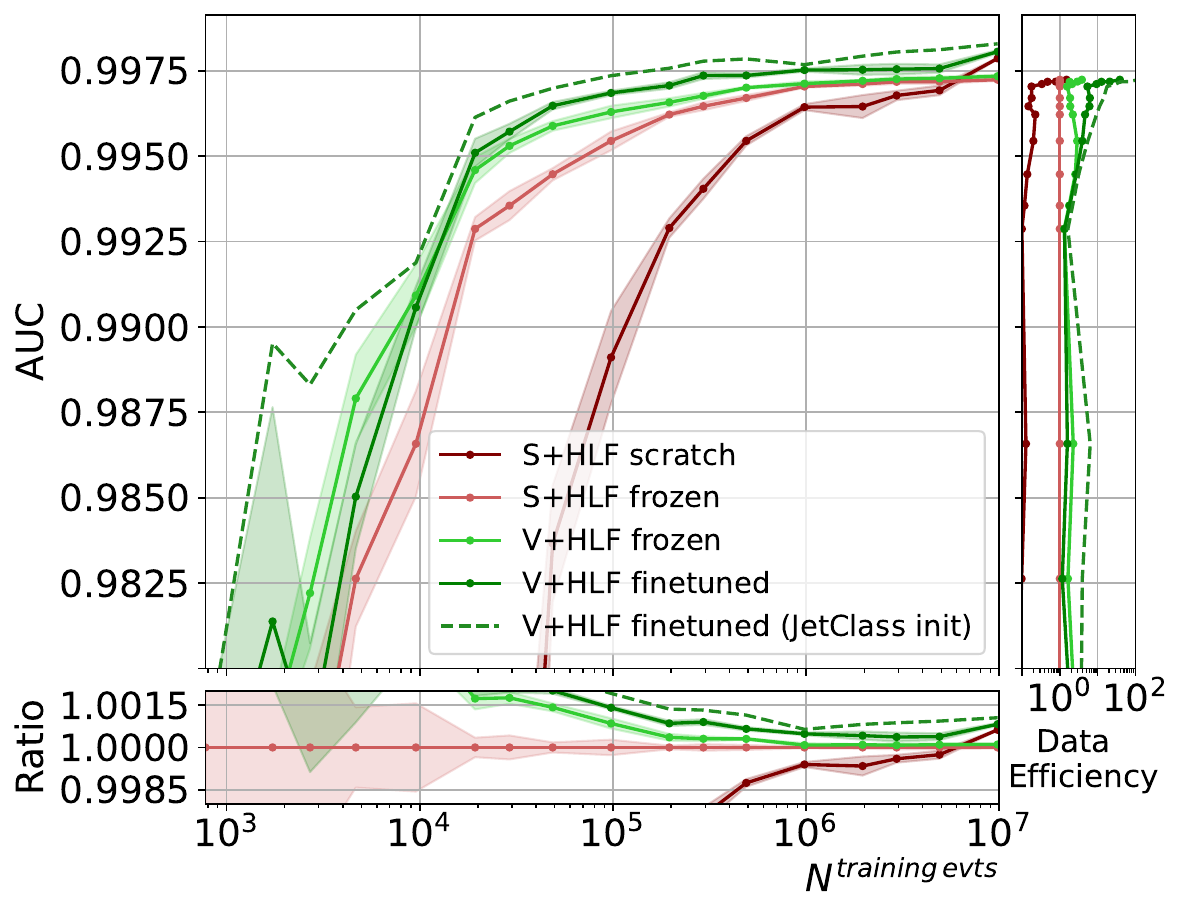}
    \caption{AUC performance as a function of labeled downstream examples. Methods from foundation models such as large-scale pretraining, finetuning, high-dimensional embedding yield significant benefits in performance and data efficiency over the baseline (\texttt{S+HLF}).}
    \label{fig:AUC}
\end{figure}

\begin{figure}[H]
    \centering
    \includegraphics[width=0.42\textwidth]{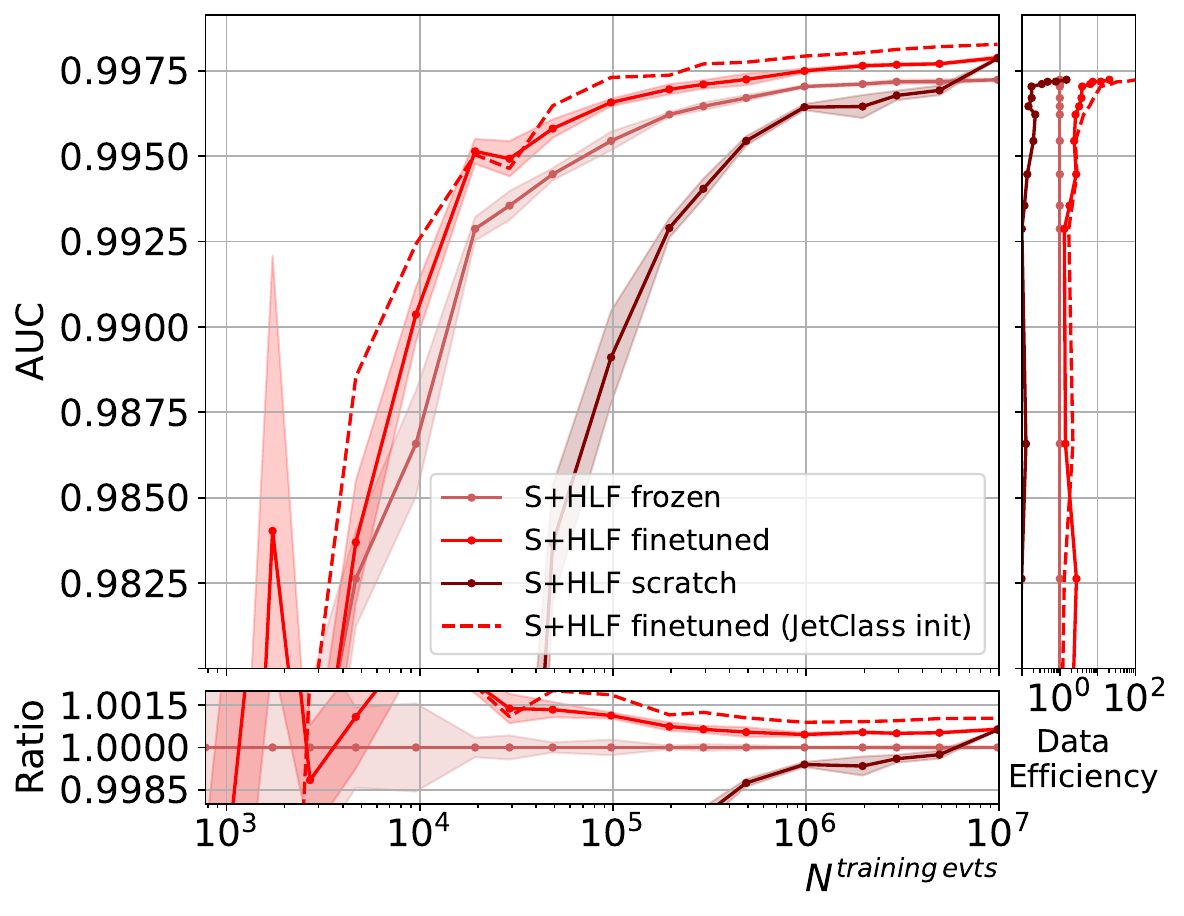}
    \includegraphics[width=0.42\textwidth]{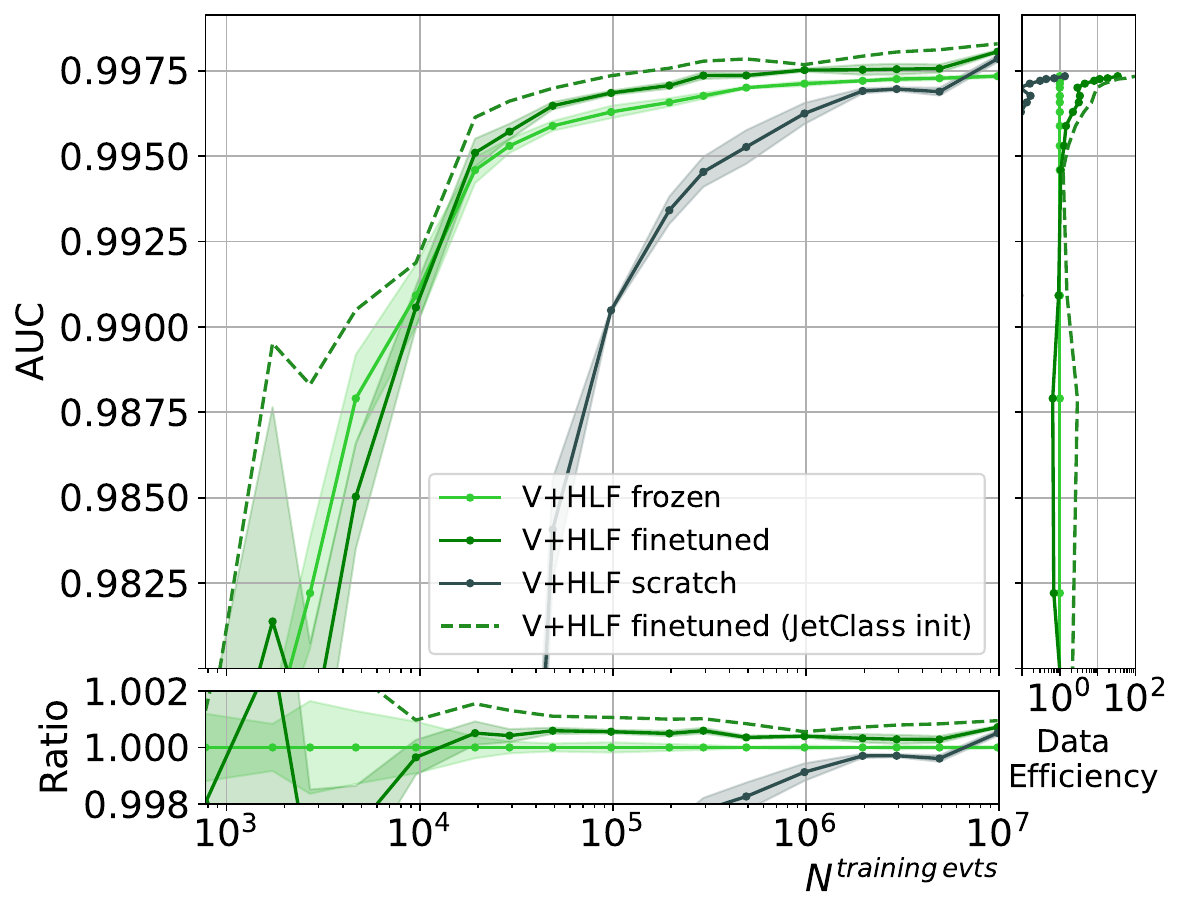}
    \includegraphics[width=0.42\textwidth]{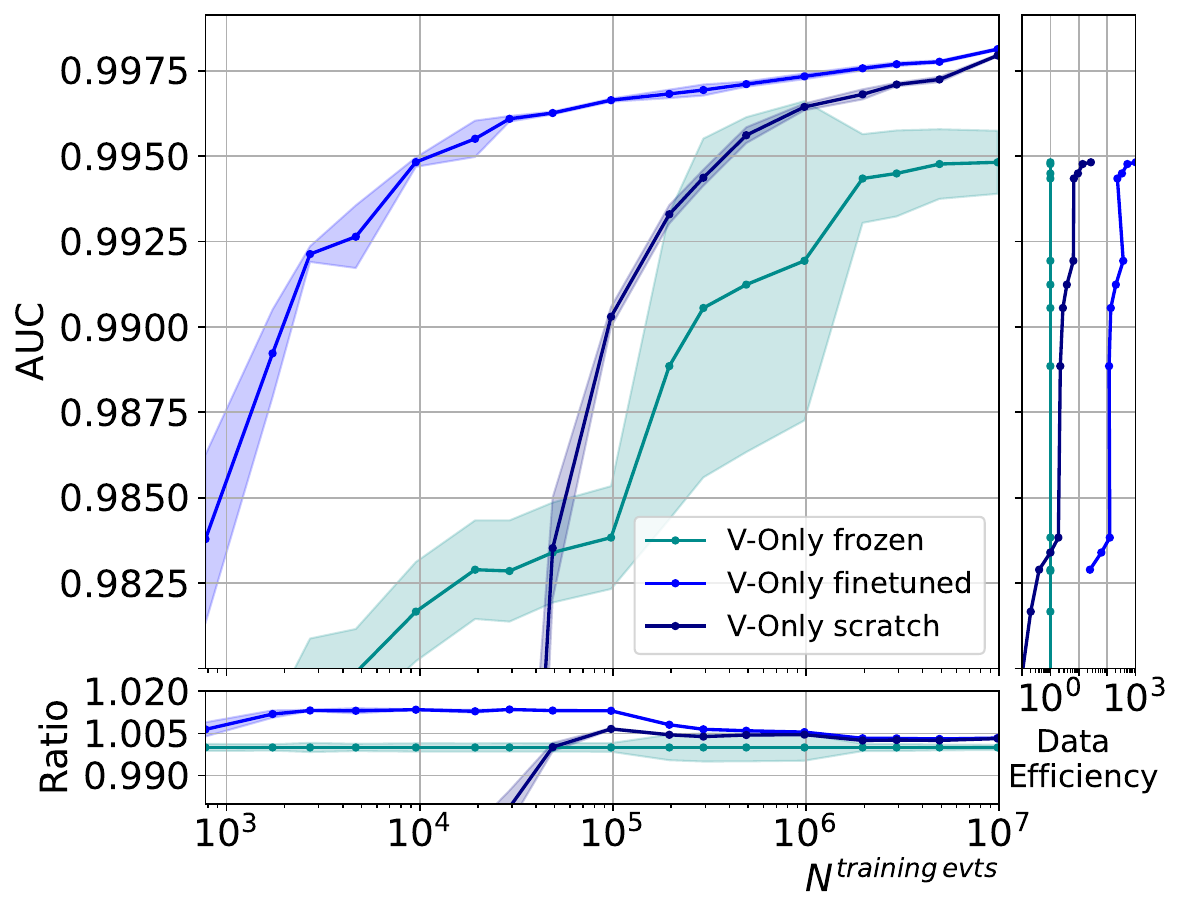}
    \caption{AUC performance as a function of labeled examples across three training strategies shown for the investigated architectures. For all architectures we see a significant benefit from finetuning over a frozen backbone. Pretraining is significantly more performant than training from scratch. For very large datasets from-scratch training can exceed a frozen backbone.}
    \label{fig:finetuning_is_good_AUC}
\end{figure}

\noindent We also show the performance of the investigated architectures in terms of Area Under the Curve (AUC) in Figures~\ref{fig:AUC},\ref{fig:finetuning_is_good_AUC}, as well as the Significance Improvement Characteristic (SIC) in Figures~\ref{fig:finetuning_is_good_SIC},\ref{fig:SIC}, which is defined as $\displaystyle \text{SIC} \equiv \frac{\epsilon_S}{\sqrt{\epsilon_B}}$, where $\epsilon_S$ and $\epsilon_B$ are the signal and background efficiencies.

% SIC

% AUC

\end{document}